\documentclass[10pt]{article}
\usepackage{array}

\usepackage{graphicx}
\usepackage{url}
\usepackage{amssymb}
\usepackage{amsmath}
\usepackage{amsthm}
\usepackage{enumitem}
\usepackage{undertilde}
\usepackage{subfigure}
\usepackage{color}      % use if color is used in text
\usepackage{multirow}
\usepackage{booktabs}
\usepackage{natbib}
\usepackage{indentfirst}
\usepackage[left=2.5cm,right=2.5cm,top=2.8cm,bottom=2.8cm,letterpaper]{geometry}
\usepackage{multirow}
\usepackage{booktabs}
\usepackage{array}
\usepackage{tabularx}
\usepackage{adjustbox}
\usepackage{bm}
\usepackage{titlesec}

\newlength{\bibitemsep}
\newlength{\bibparskip}
\let\oldthebibliography\thebibliography
\renewcommand\thebibliography[1]{%
  \oldthebibliography{#1}%
  \setlength{\parskip}{\bibitemsep}%
  \setlength{\itemsep}{\bibparskip}%
}

\titleformat{\section}
    {\centering\normalfont\Large\bfseries}{\thesection.}{1em}{}

\newlength\Origarrayrulewidth

\def\thesection{\arabic{section}}
\begin{document}

%%%%%%%%%%%%%%%%%%%%%%%%%%%%%%%%%%%%%%%%%%%%%%%%%%%%%%%%%%%%%%%%%%%%%%%%%%%%%%%%%%%%%%%%%%%%%%%%%%%%%%%%%%%%%%%%%%%%%%%%%%%%
%%%%%%%%%%%%%%%%%%%%%%%%%%%%%%%%%%%%%%%%%%%%%%%%%%%%%%%%%%%%%%%%%%%%%%%%%%%%%%%%%%%%%%%%%%%%%%%%%%%%%%%%%%%%%%%%%%%%%%%%%%%%

\renewcommand{\baselinestretch}{1.2}

\markboth{\hfill{\footnotesize\ Taewoon Kong and Brani Vidakovic} \hfill}
{\hfill {\footnotesize\rm Non-decimated Complex Wavelet Spectral Tools with Applications} \hfill}

\renewcommand{\thefootnote}{}
$\ $\par

%%%%%%%%%%%%%%%%%%%%%%%%%%%%%%%%%%%%%%%%%%%%%%%%%%%%%%%%%%%%%%%%%%%%%%%%%%%%%%%%%%%%%%%%%%%%%%%%%%%%%%%%%%%%%%%%%%%%%%%%%%%%

\fontsize{10.95}{14pt plus.8pt minus .6pt}\selectfont
\vspace{0.8pc}
\centerline{\large\bf Non-decimated Complex Wavelet Spectral Tools with Applications}
\vspace{2pt}
\centerline{\large\bf }
\vspace{.4cm}
\centerline{Taewoon Kong$^*$,  \url{twkong@gatech.edu}}
\centerline{Brani Vidakovic, \url{brani@gatech.edu}}
\vspace{.2cm}
\centerline{\it H. Milton Stewart School of Industrial \& Systems Engineering}
\centerline{\it Georgia Institute of Technology, Atlanta, USA}
\vspace{.55cm}
\fontsize{9}{11.5pt plus.8pt minus .6pt}\selectfont

\begin{abstract}

 In this paper we propose spectral tools based on non-decimated complex wavelet transforms implemented by their matrix formulation. This non-decimated complex wavelet spectra utilizes both real and imaginary parts of complex-valued wavelet coefficients via their modulus and phases. A structural redundancy in non-decimated wavelets and a componential redundancy in complex wavelets act in a synergy when extracting wavelet-based informative descriptors. In particular, we suggest an improved way of separating signals and images based on their scaling indices in terms of spectral slopes and information contained in the phase in order to improve performance of classification. We show that performance of the proposed method is significantly improved when compared with procedures based on standard versions of wavelet transforms or on real-valued wavelets.
 It is worth mentioning that the matrix-based non-decimated wavelet transform can handle signals of an arbitrary size and in 2-D case, rectangular images of possibly different and non-dyadic dimensions. This is in contrast to the standard wavelet transforms where algorithms for handling objects of non-dyadic dimensions requires either data preprocessing or customized algorithm adjustments.

 To demonstrate the use of defined spectral methodology we provide two examples of application on real-data problems: classification of visual acuity using scaling in pupil diameter dynamic in time and diagnostic and classification of digital mammogram images using the fractality of digitized images of the background tissue. The proposed tools are contrasted with the traditional wavelet based counterparts.

\vspace{9pt}
\par
\end{abstract}\par

\fontsize{10.95}{14pt plus.8pt minus .6pt}\selectfont

%\newpage
%
%\tableofcontents
%
%\newpage

%------------------------------------------
\section{Introduction}{\label{sec-Intro}}
Wavelets have become standard tools in signal and image processing. Of many versions of a wavelet transforms that are used in such applications, a popular version is a complex wavelet transform. We denote it as $\text{WT}_\text{\large{c}}$ where c refers to complex instead of CWT that usually stands for continuous wavelet transform.
%Symmetric Daubechies Wavelets produce redundant complex-valued wavelet coefficients through a multiresolution representation of real signals.
In the past, the multiresolution analysis based on the complex-valued coefficients had not been widely utilized since the resulting redundant representations of real signals seemed to be uninformative \citep{Lina1997}.
%However, it has received much attention these days because of numerous desirable properties compared to other versions of a wavelet transform.
It is agreed among experts that desirable properties for basis functions in functional representation of signals and images should be orthogonality, symmetry, and compact support \citep{Gao2010}.
Orthogonality is important because of representational parsimony \citep{Mallat2009}. In particular, the orthogonality is important for a coherent definition of power spectra because of energy preservation.
The symmetry is especially desired when dealing with images \citep{Antonini1992}.
In particular, the study in \citet{Simoncelli1996} showed that symmetric basis functions can prevent directional distortions via an orientation-free representation of features.
Finally, functional representations should be computationally efficient and local which requires compact support for decomposing functions.
These three desirable properties in the wavelet context are only available by the orthogonal complex wavelets with an odd number of vanishing moments.
The Haar wavelet is an exception \citep{Lawton1993}.
For Daubechies complex wavelets in \citet{Lina1997}, these characteristics result from the underlying differential operators defining the complex-valued multiresolution.
Even though the complex wavelets are orthogonal, the representations are redundant because of complex-valued coefficients.
This provides for a potential benefit of phase information \citep{Jeon2014}.
Because of this supplemental phase information the complex wavelets have been utilized in various fields including motion estimation \citep{Magarey1998}, texture image modeling \citep{Portilla2000}, signal denoising \citep{Achim2005, Remenyi2014}, NMR spectra classification \citep{Kim2008}, and mammogram images classification \citep{Jeon2014}.

Although orthogonal transforms are minimal, mathematically elegant, and easy to implement, they suffer from the Balian-Low obstacle concerning simultaneous locality in the time and scale domains.
Redundant dictionaries can be constructed that preserve the ease of computation and do not suffer from the Balian-Low limitations by sacrificing the orthogonality property.
As a compromise, the non-decimated wavelet transform (NDWT) is a superposition of many orthogonal transforms, and as such preserves the ease of computation but results in redundant representations.
As we look at some alternative names of NDWT such as ``stationary wavelet transform,'' ``time-invariant wavelet transform,'' ``{\it \'a trous} transform,'' or ``maximal overlap wavelet transform'', they all refer to its two
properties: redundancy and translation invariance, both absent in traditional orthogonal discrete wavelet transforms (WT).
%Even if some researches on the names use somewhat different terms, they are inherently equivalent to the NDWT.
Non-decimated wavelet transform represents a dense discrete sample of coefficients from continuous wavelet transforms, which results in their structural redundancy.
Operationally, non-decimated wavelet transform is performed by Mallat's algorithm without decimation: a repeated filtering with a minimal shift at all dyadic scales.
Consequently, at each multiresolution level the number of wavelet coefficients is the same as the size of the original data.
Although the non-decimated wavelet transform increases computational complexity, it has been widely used particularly because of the usefulness of redundancy and an easy way to adjust for the energy preservation.
More details on some additional benefits over the standard WT can be found in \citet{Kang2016}.

In this paper we propose a non-decimated complex wavelet transform ($\text{NDWT}_\text{\large{c}}$) that is a combination of the aforementioned two types of wavelet transform. The $\text{WT}_\text{\large{c}}$ produces complex-valued redundant type of wavelet coefficients and the non-decimated wavelets have a redundant structure of wavelet coefficients. We call the former componential redundancy and the latter structural redundancy. Since they represent different types of redundancy, we suggested that their combination can be beneficial in feature extraction.

%%% first novelty %%%
A study in \citet{Jeon2014} suggested a classification procedure for mammogram images based on obtained spectral slope based on the modulus and average of phases at the finest level, constructed from coefficients in $\text{WT}_\text{\large{c}}$.
The novelty of that approach was that it calculated a descriptor based on phases of complex wavelet coefficients and used it as an additional input in machine learning tasks.
The authors in \citet{Jeon2014} showed that use of phase increased the precision of the classification.

We suggest that this performance can be improved by incorporating the phase information from all detail levels in the multiresolution analysis.
Different levels of detail in the multiresolution hierarchy carry almost independent information on the signal behavior on various scales.
Experimental evidence showed that phase information from the coarser scales can serve as useful summaries in classification algorithms.
Besides, the accuracy can be increased more when the $\text{WT}_\text{\large{c}}$ is used with the NDWT together because of the redundancy.
This is because level based summaries are obtained from large number of coefficients.
One criticism could be that the increased dependence of the coefficients within the level in non-decimated transforms can be detrimental to the summary statistics.
It is true that this would be an impediment for the estimation inference, but not so for the classification because the possible bias in the summaries affects the coefficients from different classes in the same way.

The one of disadvantages of standard wavelet transforms is that they are efficiently applied only to signals and square-sized images whose dimensions are dyadic, even for the complex wavelets and convolution-based non-decimated wavelets \citep{Lina1999, Percival2000}.
In practice, this is a serious limitation and to overcome it one increases the computational complexity. We construct the matrix-based NDWT in \citet{Kang2016} with complex-valued filters in order to have an automatic transform for the signals and images of arbitrary size. Thus, this property of matrix-based implementation gives us more flexibility that is necessary for tackling real-world data.
We note that the use of matrix-based transform is not practical for very long 1-D signals, in which case special sparse matrix representation and operations have to be used, which ultimately boils down to the Mallat's algorithm.
But for the 2-D transforms, this is not the case. If the computer can store the data matrix, then it can store the transformation matrix as well and can perform the matrix multiplication to transform.
Most real-life images are of order of tens megapixels, so the matrix-type transforms are readily implementable even on modest personal computers.

The paper is organized as follows.
Section \ref{sec-NDCWT} describes the $\text{NDWT}_\text{\large{c}}$ for 1-D and 2-D cases, respectively. For the 2-D case, we present a scale-mixing 2-D $\text{NDWT}_\text{\large{c}}$.
Section \ref{sec-NDCwavespec} illustrates a non-decimated complex wavelet spectra based on the modulus of the wavelet coefficients, while Section \ref{sec-phase} proposes an effective way of utilizing the phase information leading to phase-based summaries enhancing discriminatory analysis of signal and images.
Section \ref{sec-app} demonstrates a power of the proposed method with 1-D and 2-D applications and Section \ref{sec-Conc} contains some remarks and directions for future study.

\section{Non-decimated Complex Wavelet Transform}{\label{sec-NDCWT}}
The wavelet and scaling functions for complex wavelets in \citet{Lawton1993}, \citet{Strang1996}, \citet{Lina1999}, and \citet{Zhang1999} satisfy
\begin{equation}\label{compfunc1}
\phi(x) = \sum_{k \in \mathbb{Z}} h_k \sqrt{2} \phi(2x-k) = h(x) + i g(x),
\end{equation}
\begin{equation}\label{compfunc2}
\psi(x) = \sum_{k \in \mathbb{Z}} g_k \sqrt{2} \phi(2x-k) = w(x) + i v(x),
\end{equation}
where $h_k$ denotes the low pass filter and $g_k$ is defined as
\begin{equation*}\label{relation}
g_k = (-1)^k \overline{h_{1-k}},
\end{equation*}
where $\overline{h_{1-k}}$ denotes a complex conjugate of $h_{1-k}$.

Using the complex wavelet bases, in this section, we define the non-decimated complex wavelet transform ($\text{NDWT}_\text{\large{c}}$) separately for 1-D and 2-D cases by connecting the complex scaling and wavelet
functions in non-decimated fashion.

\subsection{1-D case}{\label{sec-1dNDCWT}}
Suppose that a data vector $\bm{\mathit{y}} = (y_0, y_1, \dots, y_{m-1}$) of size $m$ is given and that a multiresolution framework is specified. To understand the interplay between transform applied
to discrete data and wavelet series representation of the function, we can link the data vector $\bm{\mathit{y}}$ to a function $f$ in terms of shifts of the scaling function at a multiresolution level $J$ as follows:
\begin{eqnarray*}
f(x) &=& \sum_{k=0}^{m-1} y_k \phi_{J,k}(x)
\end{eqnarray*}
where $J-1 < \log_2m  \le  J$, i.e. $J = \lceil \log_2 m \rceil$, and
\begin{gather*}
\phi_{J,k}(x) = 2^{\frac{J}{2}} \phi(2^J (x - k) ).
\end{gather*}
Since we consider the complex-valued filters in this wavelet transform, the scaling function is also complex-valued function as in Equation (\ref{compfunc1}). Note that $2^J (x - k)$ is used as an argument of scaling function, instead of $2^J x - k$ as in traditional wavelet transform, since we do not decimate.

Similarly, we can also express the data interpolating function $f$ in terms of wavelet coefficients as
\begin{eqnarray*}\label{NDCfx}
f(x)&=& \sum_{k=0}^{m-1} c_{J_0,k} \phi_{J_0,k}(x) + \sum_{j=J_0}^{J-1} \sum_{k=0}^{m-1}  d_{j,k}\psi_{j,k}(x)
\end{eqnarray*}
where
\begin{eqnarray*}\label{NDCref}
\phi_{J_0,k}(x) &=& 2^{\frac{J_0}{2}} \phi\left( 2^{J_0} (x - k) \right),\\
\psi_{j,k}(x) &=& 2^{\frac{j}{2}}  \psi\left(2^{j} (x - k) \right), \nonumber
\end{eqnarray*}
and $J_0$ is the coarsest decomposition level.
Note that the non-decimated complex wavelet coefficients, $c_{J_0,k}$ and $d_{j,k}$, have both real and imaginary parts as
\begin{eqnarray}\label{NDCcoeffi}
c_{J_0,k} &=& \text{Re}(c_{J_0,k}) + i \cdot \text{Im}(c_{J_0,k}), \nonumber\\
d_{j,k} &=& \text{Re}(d_{j,k}) \hspace{0.16cm} + i \cdot \text{Im}(d_{j,k}) \;\; \mbox{for} \;\; j=J_0, \dots, J-1.
\end{eqnarray}
On the basis of these complex-valued wavelet coefficients we will, in the later sections, construct a wavelet spectra of modulus and as well as level-dependent phase summaries.

For a decomposition depth $p=J-J_0$, the $\text{NDWT}_\text{\large{c}}$ transform of a vector $\bm{\mathit{y}}$ consists of
a vector of  ``smooth'' coefficients  serving as a coarse approximation of $\bm{\mathit{y}}$,
\begin{equation*}
\bm{\mathit{c}}_{(J_0)} = (c_{J_0,0}, c_{J_0,1}, \dots, c_{J_0,m-1}),
\end{equation*}
and a set of ``detail'' coefficients containing information about the localized features in the data
\begin{equation*}
\bm{\mathit{d}}_{(j)} = ( d_{j,0}, d_{j,1}, \dots, d_{j,m-1}), \;\; j=J_0, \dots, J-1.
\end{equation*}

\begin{figure}[!ht]
\centering
\includegraphics [scale=0.5]{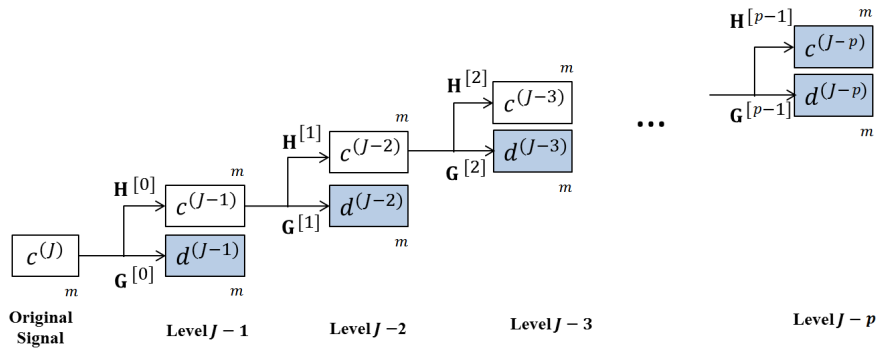}
\caption{Graphical illustration of the $\text{NDWT}_\text{\large{c}}$ Mallat algorithm. The $\text{NDWT}_\text{\large{c}}$ decomposes the original signal of size $m$ to $p+1$ multiresolution subspaces including $p$ levels of detail coefficients and one level of coarse  coefficients. The coefficients of the transform  ${\bm d}^{(J-1)}, \bm{d}^{(J-2)}, \dots,  \bm{d}^{(J-p)},$ and $\bm c^{(J-p)}$ are in the shaded boxes.}
\label{fig:NDWTG}
\end{figure}

The total number of coefficients of each vector is always $m$,
which implies the redundancy of non-decimated transforms in contrast to the length-preserving standard WT.
This results in total of $(p+1) \times m$  wavelet coefficients, with $p$ standing for number of levels of detail and 1 for the coarse level.
The constancy of the level-wise shifts enables the $\text{NDWT}_\text{\large{c}}$ to be time invariant.
The Mallat type algorithm for $\text{NDWT}_\text{\large{c}}$  is graphically illustrated in Figure \ref{fig:NDWTG}.
The coefficients in shaded boxes comprise the transform.

Since the non-decimated wavelet transform is linear, the wavelet coefficients can be linked to the original signal by a matrix multiplication.
For the proposed $\text{NDWT}_\text{\large{c}}$, we apply the complex scaling and wavelet filters in Equation (\ref{compfunc1}) and (\ref{compfunc2}) into the matrix formulation of NDWT defined in \citet{Kang2016} to obtain a matrix $W_m^{(p)}.$
This matrix corresponds to a non-decimated complex wavelet transform of depth $p$, that is with $p$  levels of detail, and with $m$ as the size of input data.
As we indicated earlier, the reason why we prefer the matrix-formulation is that it provides more flexibility especially in the 2-D case, with only a slight increase of computational complexity.
Details for constructing the $W_m^{(p)}$ can be found in \citet{Kang2016}.
With use of $W_m^{(p)}$ we can transform a 1-D signal $\boldsymbol{y}$ of size $m$ to a non-decimated complex vector $\boldsymbol{d}$
\begin{eqnarray*}\label{eq:correctconst}
\boldsymbol{d}&=& {W}_{m}^{(p)} \cdot \boldsymbol{y}
\end{eqnarray*}
where  $p$ is a depth of the transform and $p$ and $m$ are arbitrary.
When the matrix wavelet transforms is used, one needs a weight matrix, ${T}_{m}^{(p)}$, to reconstruct back $\boldsymbol{y}$ from $\boldsymbol{d}$.
The need for a weight matrix is caused by the inherent redundancy of the transform, and serves for deflation of the energy inflated by the transform.
The weight matrix ${T}_{m}^{(p)}$ is defined as
\begin{equation}\label{eq:1dweight}
 {T}_{m}^{(p)}=\mbox{diag}(\overbrace{1/2^p, \dots,1/2^{p}}^\text{$2m$},\overbrace{1/2^{p-1},\dots,1/2^{p-1}}^\text{$ m$},\dots, \overbrace{1/2, \dots,1/2}^\text{$m$}).
\end{equation}
By using the weight matrix, the perfect reconstruction can be obtained as
\begin{eqnarray*}\label{eq:correctreconst}
\boldsymbol{y} &=&  ({W}_{m}^{(p)})' \cdot {T}_{m}^{(p)} \cdot \boldsymbol{d}.
\end{eqnarray*}

\subsection{2-D case}{\label{sec-2dNDCWT}}

Next, we extend the 1-D definitions to the scale-mixing 2-D $\text{NDWT}_\text{\large{c}}$ of $f(x,y)$ where $(x,y) \in  \mathbb{R}^2$. The representation of non-decimated complex wavelets in 2-D can be implemented through one scaling function and three wavelet functions defined using Equations (\ref{compfunc1}) and (\ref{compfunc2}) as follows:
\begin{eqnarray}\label{eq:2dsNDCwavscafunctions}
\phi(x,y) &=& \phi(x) \phi(y) = \Theta(x,y) + i  \Psi(x,y), \nonumber\\
\psi^{(h)}(x,y) &=& \phi(x) \psi(y) = \xi^{(h)}(x,y) + i \zeta^{(h)}(x,y), \nonumber\\
\psi^{(v)}(x,y) &=& \psi(x) \phi(y) = \xi^{(v)}(x,y) + i \zeta^{(v)}(x,y),  \nonumber\\
\psi^{(d)}(x,y) &=& \psi(x) \psi(y) = \xi^{(d)}(x,y) + i \zeta^{(d)}(x,y),
\end{eqnarray}
where symbols $h,v,$ and $d$ denote the horizontal, vertical, and diagonal directions, respectively.
This $h,v,d$ -notation is standardly used in 2-D wavelet literature and refers to directions in which the
features are located in the hierarchy of multiresolution subspaces.
%Thus, correspondingly we can define
%\begin{eqnarray*}
%\Theta_{j_1,j_2,k_1,k_2}(x,y) &=& h_{j_1,k_1}(x)h_{j_2,k_2}(y) - g_{j_1,k_1}(x)g_{j_2,k_2}(y), \\
%\Psi_{j_1,j_2,k_1,k_2}(x,y) &=& h_{j_1,k_1}(x)g_{j_2,k_2}(y) + g_{j_1,k_1}(x)h_{j_2,k_2}(y), \\
%\xi^{(h)}_{j_1,j_2,k_1,k_2}(x,y) &=& h_{j_1,k_1}(x)w_{j_2,k_2}(y) - g_{j_1,k_1}(x)v_{j_2,k_2}(y), \\
%\zeta^{(h)}_{j_1,j_2,k_1,k_2}(x,y) &=& h_{j_1,k_1}(x)v_{j_2,k_2}(y) + g_{j_1,k_1}(x)w_{j_2,k_2}(y), \\
%\xi^{(v)}_{j_1,j_2,k_1,k_2}(x,y) &=& w_{j_1,k_1}(x)h_{j_2,k_2}(y) - v_{j_1,k_1}(x)g_{j_2,k_2}(y), \\
%\zeta^{(v)}_{j_1,j_2,k_1,k_2}(x,y) &=& w_{j_1,k_1}(x)g_{j_2,k_2}(y) + v_{j_1,k_1}(x)h_{j_2,k_2}(y), \\
%\xi^{(d)}_{j_1,j_2,k_1,k_2}(x,y) &=& w_{j_1,k_1}(x)w_{j_2,k_2}(y) - v_{j_1,k_1}(x)v_{j_2,k_2}(y), \\
%\zeta^{(d)}_{j_1,j_2,k_1,k_2}(x,y) &=& w_{j_1,k_1}(x)v_{j_2,k_2}(y) + v_{j_1,k_1}(x)w_{j_2,k_2}(y), \\
%\end{eqnarray*}

\subsubsection{Scale-Mixing 2-D Non-decimated Complex Wavelet Transform}{\label{sec-sm2dNDCWT}}
Although various versions of the 2-D WT can be constructed by appropriate tessellations of the detail spaces, here we utilize the scale-mixing 2-D $\text{NDWT}_\text{\large{c}}$.
As we will argue later, the use scale-mixing version is motivated by its remarkable flexibility, compressibility, and ease of computation.

For the scale-mixing 2-D $\text{NDWT}_\text{\large{c}}$, we define the wavelet atoms as follows:
\begin{eqnarray}\label{eq:2dscalemixtransNDCfunctions}
\phi_{J_{01},J_{02},k_1,k_2}(x,y) &=& \Theta_{J_{01},k_1,k_2}(x,y) + i \Psi_{J_{02},k_1,k_2}(x,y), \nonumber \\
\psi_{J_{01},j_2,k_1,k_2}^{(h)}(x,y) &=& \xi^{(h)}_{J_{01},k_1,k_2}(x,y) + i \zeta^{(h)}_{j_2,k_1,k_2}(x,y), \nonumber \\
\psi_{j_1,J_{02},k_1,k_2}^{(v)}(x,y) &=& \xi^{(v)}_{j_1,k_1,k_2}(x,y) + i \zeta^{(v)}_{J_{02},k_1,k_2}(x,y), \nonumber \\
\psi_{j_1,j_2,k_1,k_2}^{(d)}(x,y) &=& \xi^{(d)}_{j_1,k_1,k_2}(x,y) + i \zeta^{(d)}_{j_2,k_1,k_2}(x,y),
\end{eqnarray}
where $k_1 = 0, \dots, m-1$, $k_2 = 0, \dots, n-1$, $j_1 = J_{01}, \dots, J-1$, $j_2 = J_{02}, \dots, J-1$, and $J = \lceil \log_2 \min(m,n) \rceil$. Note that $J_{01}$ and $J_{02}$ are the coarsest decomposition levels of rows and columns. Then any function $f \in L_2(\mathbb{R}^2)$ can be expressed as
\begin{eqnarray*}\label{eq:2dscalemixtransNDC*}
f(x, y) &= & \sum_{k_1}  \sum_{k_2} c_{J_{01}, J_{02}, k_1,k_2} \phi_{J_{01}, J_{02}, k_1,k_2}(x,y)  \nonumber \\
&+ &  \sum_{j_2>J_{02}} \sum_{k_1}  \sum_{k_2} d_{J_{01}, j_2, k_1,k_2}^{(h)} \psi_{J_{01}, j_2, k_1,k_2}^{(h)}(x,y) \nonumber\\
&+ &  \sum_{j_1>J_{01}} \sum_{k_1}  \sum_{k_2} d_{j_1, J_{02}, k_1,k_2}^{(v)} \psi_{j_1, J_{02}, k_1,k_2}^{(v)}(x,y) \nonumber \\
&+ & \sum_{j_1>J_{02}} \sum_{j_2>J_{01}} \sum_{k_1}  \sum_{k_2} d_{j_1, j_2, k_1,k_2}^{(d)} \psi_{j_1, j_2, k_1,k_2}^{(d)}(x,y),
\end{eqnarray*}
which defines a scale-mixing $\text{NDWT}_\text{\large{c}}$.
Unlike the standard 2-D $\text{NDWT}_\text{\large{c}}$ denoting a scale as only $j$, we denote such mixed two scales as a pair $(j_1, j_2)$ capturing the energy flux between the scales.

Finally, the resulting scale-mixing non-decimated complex wavelet coefficients are
\begin{eqnarray}
c_{ J_{01},J_{02}, k_1, k_2 } &=&  2^{\frac{J_{01}+J_{02}}{2}} \iint   f(x,y) \overline{\phi}_{J_{01},J_{02}, k_1, k_2} (x, y)\; dxdy \nonumber \\
&=& \text{Re}(c_{J_{01}, J_{02}, k_1,k_2}) + i \cdot \text{Im}(c_{J_{01}, J_{02}, k_1,k_2}), \nonumber\\
d_{ J_{01},j_{2},k_1, k_2 }^{(h)} &=& 2^{\frac{J_{01}+j_2}{2}} \iint   f(x,y) \overline{\psi}_{J_{01},j_{2}, k_1, k_2}^{(h)}(x, y)\; dxdy \nonumber\\
&=& \text{Re}(d_{J_{01}, j_2, k_1,k_2}^{(h)}) + i \cdot \text{Im}(d_{J_{01}, j_2, k_1,k_2}^{(h)}), \\
d_{ j_{1},J_{02},k_1, k_2 }^{(v)} &=& 2^{\frac{j_1+J_{02}}{2}} \iint  f(x,y) \overline{\psi}_{j_{1},J_{02}, k_1, k_2}^{(v)}(x, y)  \; dxdy \nonumber \\
&=& \text{Re}(d_{j_1, J_{02}, k_1,k_2}^{(v)}) + i \cdot \text{Im}(d_{j_1, J_{02}, k_1,k_2}^{(v)}), \nonumber\\
d_{ j_1,j_2,k_1, k_2}^{(d)} &=&  2^{\frac{j_1+j_2}{2}} \iint   f(x,y) \overline{\psi}_{j_1,j_2, k_1, k_2}^{(d)}(x, y) \; dxdy \nonumber \\
&=& \text{Re}(d_{j_1, j_2, k_1,k_2}^{(d)}) + i \cdot \text{Im}(d_{j_1, j_2, k_1,k_2}^{(d)}). \nonumber
\label{eq:2dscalemixtransNDCcoef}
\end{eqnarray}
where $\overline{\phi}$ denotes the complex conjugate of $\phi$.
Note that the non-decimated complex wavelet coefficients in Equation (\ref{eq:2dscalemixtransNDCcoef}) have both real and imaginary parts as complex numbers.

Similar to the 1-D case, we can connect the 2-D wavelet coefficients to the original image through a matrix equation. Here we apply the complex scaling and wavelet filters in Equation (\ref{eq:2dsNDCwavscafunctions}) into the matrix formulation of NDWT to obtain $W_m^{(p_1)}$ and $W_n^{(p_2)}$ that are non-decimated complex wavelet matrices with $p_1$, $p_2$ detail levels and $m$, $n$ size of row and column, respectively. For 2-D case, using the matrix-formulation allows to use any non-square image. More rigorous details on these matrix formulation for real-valued wavelets can be found in \citet{Kang2016}.

Next, we can transform a 2-D image $\boldsymbol{A}$ of size $m \times n$ to a non-decimated complex wavelet transformed matrix $\boldsymbol{B}$ with depth $p_1$ and $p_2$ as
\begin{equation*}\label{eq:2dcorrectconst}
\boldsymbol{B}=   {W}_{m}^{(p_1)}  \cdot \boldsymbol{A} \cdot ({W}_{n}^{(p_2)})^{\dagger}
\end{equation*}
where $p_1,p_2,m$, and $n$ are arbitrary. The $W^{\dagger}$ denotes a Hermitian transpose of matrix $W$. Note that  Equation (\ref{eq:2dcorrectconst}) represents a finite-dimensional implementation of Equation (\ref{eq:2dscalemixtransNDCcoef}) for $f(x)$ sampled in a form of matrix, as $f(x,y)$.
Then the resulting transformed matrix $\boldsymbol{B}$ has a size of $(p_1 + 1) m \times (p_2 + 1) n$.
Similar to the 1-D case, for perfect reconstruction of $\boldsymbol{A}$, we need two weight matrices, that is, $p_1$- and $p_2$-level weight matrices  ${T}_{m}^{(p_1)}$ and ${T}_{n}^{(p_2)}$.
The matrices are defined as in Equation (\ref{eq:1dweight}) with different $m, n, p_1,$ and $p_2$.
By using the weight matrices, the perfect reconstruction can be performed as
\begin{equation*}\label{eq:2dcorrectreconst}
\boldsymbol{A} =  {W}_{m}^{(p_1)} \cdot  {T}_{m}^{(p_1)} \cdot  \boldsymbol{B} \cdot  {T}_{n}^{(p_2)}\cdot  ({W}_{n}^{(p_2)})^{\dagger}.
\end{equation*}

\section{Non-decimated Complex Wavelet Spectra}{\label{sec-NDCwavespec}}
High-frequency, time series data from various sources often possess hidden patterns that reveal the effects of underlying functional differences. Such patterns cannot be elucidated by basic descriptive statistics or trends in some real-life situations. For example, the high-frequency pupillary response behavior (PRB) data collected during computer-based interaction captures the changes in pupil diameter in response to various stimuli. Researchers found that there may be underlying unique patterns hidden within PRB data, and these patterns may reveal the intrinsic individual differences in cognitive, sensory and motor functions \citep{Moloney2006}. Yet, such patterns cannot be explained by the trends and traditional statistical summaries, for the magnitude of the pupil diameter depends on the ambient light, not on the inherent eye function or link to the cognitive task. When the intrinsic individual functional differences cannot be modeled by statistical tools in the domain of the data acquisition,
the transformed time/scale or time/frequency domains may help. High frequency data as a rule scale, and this scaling can be quantified by the Hurst exponent as an optional measure to characterize the patients.

The Hurst exponent is an informative summary of the behavior of self-similar processes and is also related to the presence of long memory and degree of fractality in signals and images. Among many methods for estimating the Hurst exponent, the wavelet-based methods have shown to be particularly accurate.
The main contribution of this paper is a construction of the non-decimated complex wavelet spectra
with extension of the method into the scale-mixing 2-D non-decimated complex wavelet spectra for 2-D case,
all with the goal of assessing the Hurst exponent or its equivalent spectral slope. As a bonus, the complex valued wavelets would provide informative multiscale phase information.

Next we briefly overview the notion of self-similarity and its link with the Hurst exponent.
Suppose that a random process $\{ X(t), t \in \mathbb{R} \}$ for some  $\lambda> 0$ satisfies
\begin{equation*}
X(\lambda t) \stackrel{d}{=} \lambda^H X(t) \;\; \mbox{for any} \;\;
\end{equation*}
where $\stackrel{d}{=}$ stands for equality of all joint finite-dimensional distributions, then, $X(t)$ is self-similar with self-similarity index $H$, traditionally called Hurst exponent.

If $X(t)$ is transformed in the wavelet domain and $d_{j,k}$ is the wavelet coefficient at scale $j$ and shift $k$ in standard WT, can be shown that
\begin{equation}\label{detail11}
d_{j,k} \stackrel{d}{=} 2^{-j(H+\frac{1}{2})}d_{0,k}.
\end{equation}.
Here the notation $\stackrel{d}{=}$ denotes the equality in distribution.
For the non-decimated complex wavelets, however, $d_{j,k}$ is a complex number, as in Equation (\ref{NDCcoeffi}), and we use $|d_{j,k}|$ for a modulus of $d_{j,k}$,
\begin{equation*}
|d_{j,k}| = \sqrt{Re(d_{j,k})^2 +  Im(d_{j,k})^2}, \;\; j=J_0, \dots, J-1.
\end{equation*}
The Equation (\ref{detail11}) now can be re-stated as
\begin{equation*}\label{detail12}
|d_{j,k}| \stackrel{d}{=} 2^{-j(H+\frac{1}{2})}|d_{0,k}|, \;\; j=J_0, \dots, J-1.
\end{equation*}
If the process $X(t)$ possesses stationary increments, for any $q>0$,
 $E(|d_{0,k}|)=0$ and $E(|d_{0,k}|^q)=E(|d_{0,0}|^q)$. Thus,
\begin{equation}\label{detail2}
E(|d_{j,k}|^q) = C 2^{-jq(H+\frac{1}{2})}, \;\; j=J_0, \dots, J-1
\end{equation}
where $C=E(|d_{0,0}|^q)$.
Although $q$ could be arbitrary nonnegative, here we will use standard $q=2$  that has ``energy'' interpretation.
By taking logarithms on both sides in Equation (\ref{detail2}), we can obtain the non-decimated complex wavelet spectrum of $X(t)$ as
\begin{equation}\label{wavespectra1d}
S(j) = \log_2 (E(|d_{j,k}|^2)) = -j(2H+1) + C', \;\; j=J_0, \dots, J-1.
\end{equation}

\begin{figure}[!ht]
\centering
\includegraphics [width=\linewidth,height=\textheight,keepaspectratio]{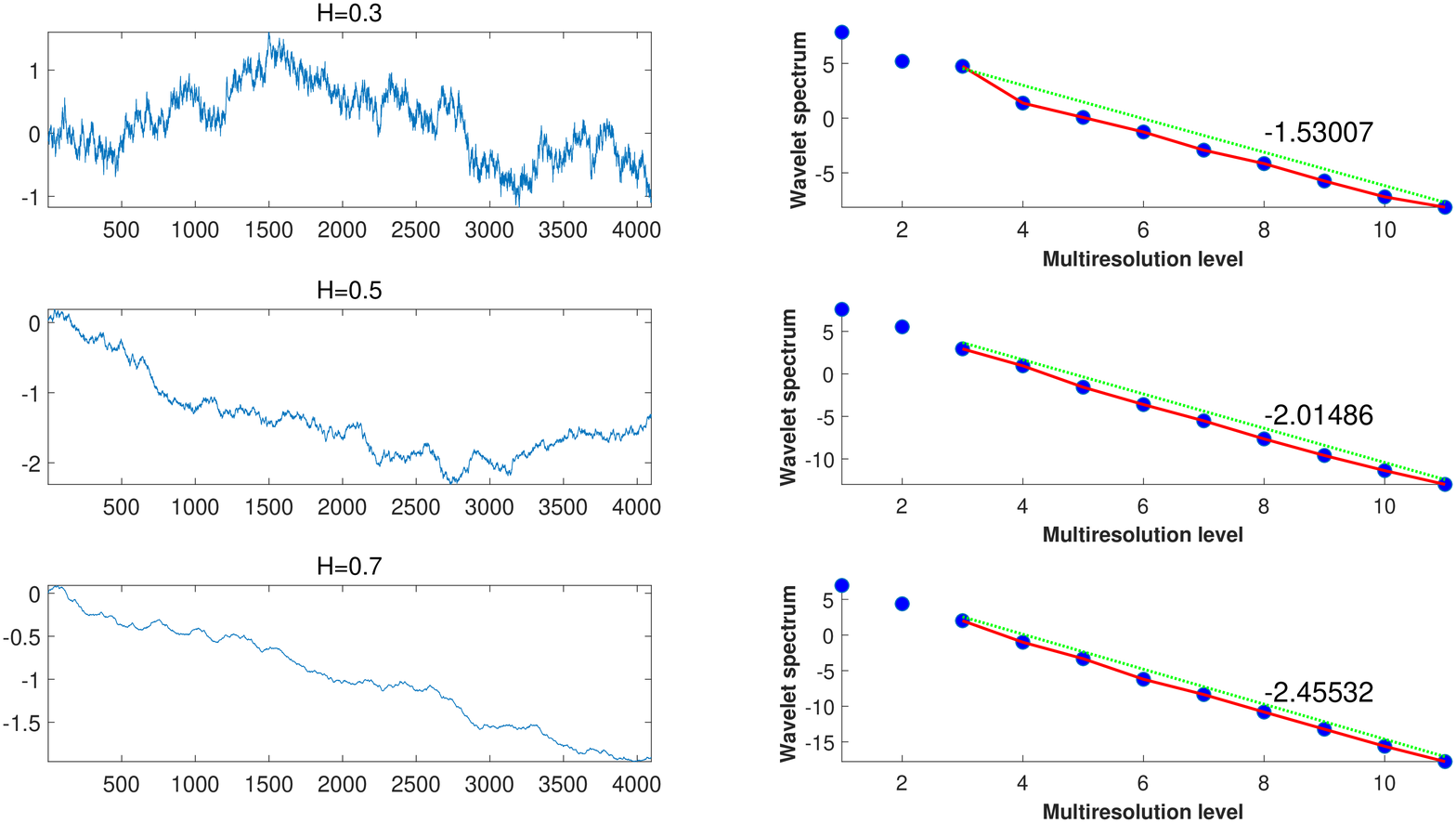}
\caption{Examples of non-decimated complex wavelet spectra using the modulus of coefficients. The slopes are -1.53007, -2.01486, and -2.45532 corresponding to estimator $\hat{H}= 0.2650, 0.5074, \; \mbox{and} \; 0.7277$. The original 4096-length signals were simulated as a fBm with Hurst exponent 0.3, 0.5, and 0.7.}
\label{exwaveletspectramod}
\end{figure}

Note that the wavelet spectrum describes the relationship between the scales and energies at the scales.
If along the scales the energies decay regularly, this indicates that there is a regular scaling in the data, and we can measure a self-similarity via a rate of energy decay.
Operationally, we find the slope in regression of log energies to scale indices, as in Equation (\ref{wavespectra1d}), and use it to estimate the Hurst exponent.
For discrete observed data of size $m$, we use empirical counterpart of $S(j)$  defined as
\begin{equation*}\label{wavespectra1dempi}
\hat{S}(j) = \log_2 \frac{1}{m} \sum_{k=1}^{m} |d_{j,k}|^2 = \log_2 \overline{|d_{j,k}|^2}, \;\; j=J_0, \dots, J-1.
\end{equation*}
We can plot the set of $\hat{S}(j)$ against $j$ as $\big( j, \; \hat{S}(j) \big),$ which is called $2$nd order Logscale Diagram (2-LD) and this is the wavelet spectra as displayed in Figure \ref{exwaveletspectramod}.
Finally, we can estimate the slope of the spectra usually by regression methodology (an ordinary, weighted, or robust regression) and use it to estimate the Hurst exponent $H$, as $\hat{H}= -(\mbox{slope} +1)/2$.
More details on wavelet spectra method and its applications can be found in \citet{Veitch1999}, \citet{Mallat2009}, \citet{Ramirez2013}, and \citet{Roberts2017}.

\subsection{Scale-Mixing 2-D Non-decimated Complex Wavelet Spectra}{\label{sec-scNDCwavespec}}
To introduce a scale-mixing 2-D non-decimated complex wavelet spectra, consider a 2-D fractional Brownian motion (fBm) in two dimensions, $B_H(\mathbf{u})$ for $\mathbf{u} \in [0,1] \times [0,1]$ and $H \in (0,1)$. The 2-D fBm, $B_H(\mathbf{u})$, is a random process with stationary zero-mean Gaussian increments leading to
\begin{equation*}
B_H(a\mathbf{t}) \stackrel{d}{=} a^H B_H(\mathbf{t}) \;\; \mbox{for any} \;\; a \geq 0.
\end{equation*}
For this process, the scale-mixing non-decimated complex wavelet detail coefficients can be defined as
\begin{equation*}\label{detailfBm}
d_{(j_1,j_2+s,k_1,k_2)}=2^{\frac{1}{2}(j_1+j_2+s)}\int B_{H}(\mathbf{u})
\overline{\psi} \left( 2^{j_1}(u_{1}-k_{1}),2^{j_2+s}(u_{2}-k_{2})  \right)  d\mathbf{u}
\end{equation*}
where $\overline{\psi}$ denotes the complex conjugate of $\psi^{(d)}$ defined in Equation (\ref{eq:2dscalemixtransNDCfunctions}).
In this paper, we only consider the main diagonal hierarchy whose 2-D scale indices coincide as $j_1 = j_2 = j$ and thus $J_{01} = J_{02} = J_0$.

Since the $d_{(j,j+s,k_1,k_2)}$ is a complex number, we need to consider its modulus
\begin{equation*}
|d_{(j,j+s,k_1,k_2)}| = \sqrt{Re(d_{(j,j+s,k_1,k_2)})^2 +  Im(d_{(j,j+s,k_1,k_2)})^2}, \;\; j=J_0, \dots, J-1.
\end{equation*}
Then average of squared modulus of the coefficients is calculated as
\begin{eqnarray}\label{mixd}
& &\mathbb{E}\left[ |d_{(j,j+s,k_1,k_2)} |^{2}\right] =
 2^{2j+s}\int \psi\left(2^j (u_{1}-k_{1}),2^{j+s}(u_{2}-k_{2}) \right) \nonumber \\
& &~~~~~~~\times \overline{\psi}\left(2^j (v_{1}-k_{1}),2^{j+s}(v_{2}-k_{2}) \right)  \mathbb {E}\left[ B_{H}(\mathbf{u})B_{H}(\mathbf{v}) \right]d\mathbf{u} \ d\mathbf{v}.
\end{eqnarray}
As a result, the Equation (\ref{mixd}) can be restated as
\begin{equation}\label{spectrumfbm}
\mathbb{E}\left[ |d_{(j,j+s,k_1,k_2)} |^{2}\right] = 2^{-j(2H+2)}\  V_{\psi,s}(H),
\end{equation}
and its proof is provided in \citet{Jeon2014}. Note that   $V_{\psi,s}(H)$ does not depend on the scale $j$ but on $\psi$, $H$ and $s$.
Finally, the scale-mixing 2-D non-decimated complex wavelet spectrum is defined by taking logarithms on both sides of the Equation (\ref{spectrumfbm}),
\begin{equation*}\label{wavespectra2d}
S(j, j+s) = \log_2 (\mathbb{E}(|d_{j,j+s,k_1,k_2}|^2)) = -j(2H+2) + C', \;\; j=J_0, \dots, J-1.
\end{equation*}
Similar to the 1-D case, its empirical counterpart is
\begin{equation*}\label{wavespectra2dempi}
\hat{S}(j, j+s) = \log_2 \frac{1}{mn} \sum_{k_1=1}^{m}\sum_{k_2=1}^{n} |d_{j,j+s,k_1,k_2}|^2 = \log_2 \overline{|d_{j,j+s,k_1,k_2}|^2}, \;\; j=J_0, \dots, J-1
\end{equation*}
where $m$ and $n$ are row and column sizes, respectively.
The way of constructing wavelet spectra goes along the lines of the construction in 1-D case, except for the expressing the Hurst exponent from the slope.
In the 2-D case $H$ is estimated as $\hat{H }= -(\mbox{slope} +2)/2.$

\section{Phase-based Statistics for Classification Analysis}{\label{sec-phase}}
In the area of Fourier representations, there is a considerable of interest about the information the phase carries about signals or images \citep{Oppenheim1981, Levi1983}.
For complex wavelet domains, there is also an interest about information related to interactions between scales and spatial symmetries contained in the phase, as investigated by \citet{Lina1997}, \citet{Lina1999}, and \citet{Jeon2014}. Therefore, it is natural to explore the role of phase in the complex-valued wavelet coefficients of signals or images.
Theoretically, it is known that the original signal can be reconstructed from the phase information only.
We briefly describe two experiments conducted in \citet{Oppenheim1981} and \citet{Jeon2014} for the Fourier and wavelet transforms, respectively.
Both experiments transformed two different images of the same size to complex-valued domains and from the coefficients obtained modulus and phases.
Then the phase information was switched and images were reconstructed from the original modulus and switched phases.
Surprisingly, both reconstructed images were more alike to the phase corresponding images, that is, the phase information dominated the modulus information.
Motivated by these experiment results, \citet{Jeon2014} proposed a way of utilizing phase information for discriminatory analysis.
They suggested a summary statistic of the phases at the finest levels and demonstrated in a particular classification task the accuracy can be improved, albeit only slightly.
This is because the phases from the finest level only were used.
Wavelet coefficients at each level, however, have slightly different information on the given data, which is the one of advantages of their multiresolution nature.
Generally, the phase information from different levels may be complementary.
If we utilize phase information on the other levels, an overall accuracy would be further improved.
In this section we propose more extensive phase-based modalities using $\text{NDWT}_\text{\large{c}}$ for signal or image classification problems to improve an overall performance.

\begin{figure}[!ht]
\centering
\includegraphics [scale=0.8 , clip]{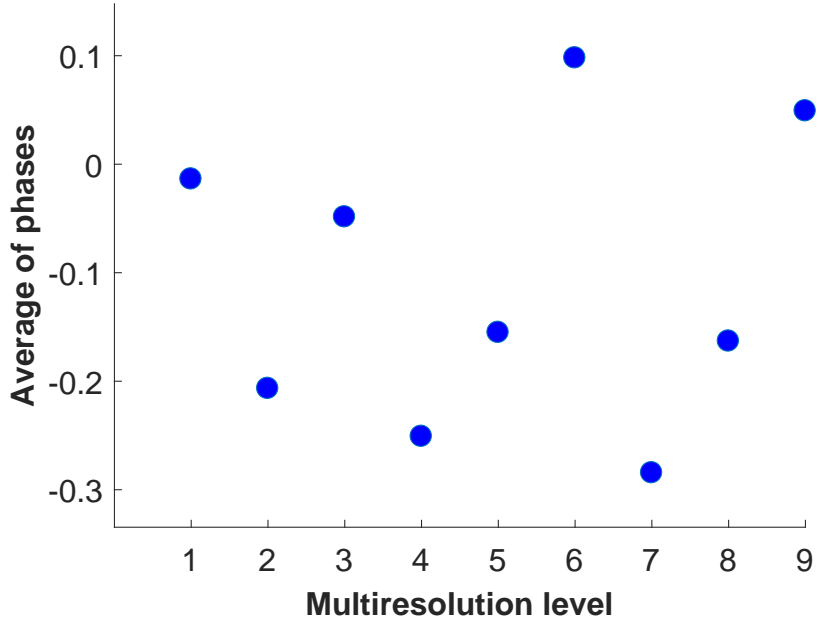}
\caption{Visualization of phase averages at all multiresolution levels.}
\label{exwaveletspectraphase}
\end{figure}

The phase of a non-decimated complex wavelet coefficient defined in Equation (\ref{NDCcoeffi}) is
\begin{eqnarray*}
\angle d_{j,k}  &=& \mbox{arctan} \Bigg( \frac{Im(d_{j,k})}{Re(d_{j,k})} \Bigg), \\
\angle d_{(j,j+s,k_1,k_2)} &=& \mbox{arctan} \Bigg(\frac{Im(d_{(j,j+s,k_1,k_2)})}{Re(d_{(j,j+s,k_1,k_2)})} \Bigg)
\end{eqnarray*}
for 1-D and 2-D cases, respectively. Then, an average of phases at level $j$ for both cases can be calculated as
\begin{eqnarray}\label{avgphase}
\angle d_j &=& \frac{1}{m} \sum_{k=1}^{m} \angle d_{j,k}, \;\; j=J_0, \dots, J-1, \\
\angle d_{j, j+s} &=& \frac{1}{mn} \sum_{k_1=1}^{m}\sum_{k_2=1}^{n} \angle d_{(j,j+s,k_1,k_2)}, \;\; j=J_0, \dots, J-1 \nonumber
\end{eqnarray}
for 1-D and 2-D cases, respectively.
Finally, we set the averages of phases at all considered multiresolution level $j$ as new descriptors in a wavelet-based classification analysis.
Note that these descriptors do not indicate any scaling regularity, unlike the modulus, as seen in Figure \ref{exwaveletspectraphase}.

\section{Applications}{\label{sec-app}}

\subsection{Application in Classifying Pupillary Signal Data}{\label{sec-pupil}}
The human computer interaction (HCI) community has been interested in evaluating and improving user performance and interaction in a variety of fields. In particular, a variety of researches have been performed to investigate the interactions of users with age-related macular degeneration (AMD) since it is one of main causes of visual impairments and blindness in people over 55 years old \citep{TSERI2002}. AMD influences high resolution vision that affects abilities of people for focus-intensive tasks such as using a computer \citep{TCSMD2002}. The research has proved that people with AMD are likely to show worse performance than ordinary people based on measures such as task times and errors on simple computer-based tasks. In this regard, mental workload due to sensory impairments is well known as a significant factor of human performance while interacting with a complicated system \citep{Gopher1986}. However, only a few studies have been performed to investigate how mental workload due to sensory impairments makes effects on the performance mentioned above. Thus, we need to consider pupil diameter that is one of significant measures of workload \citep{Loewenfeld1999, Andreassi2000}. However, the pupil has such a complex control mechanism that it is difficult to find meaningful signals from considerably noised signals of pupillary activity \citep{Barbur2003}. Therefore, it is necessary for a strong support to develop an analytical model to analyze dynamic pupil behaviors. Note that trends in high frequency of pupil-diameter measures are not significant because other factors that are not related to the pathologies could affect them, such as a change of environmental light intensity. Instead, the scaling information  can be used for the analysis since pupil-diameter measures are considered self-similar signals. Thus, we propose an analytic tool based on the wavelet spectra method described in Section \ref{sec-NDCwavespec} with phase-based modalities suggested in Section \ref{sec-phase}.

\subsubsection{Description of Data}{\label{sec-data1}}
The dataset consists of pupillary response signals for 24 subjects as described in Table \ref{pupildata}.

\begin{table}[h!tb]
\begin{center}
\begin{adjustbox}{max width=\textwidth}

\begin{tabular}{c|c|c|c|c}
  \specialrule{1.3pt}{1pt}{1pt}
   Group & N & Visual acuity & AMD & Number of samples  \\\hline \hline
  Control & 6 & 20/20 - 20/40  &  No  & 1170    \\
   Case 1 & 8 & 20/20 - 20/50 &  Yes & 1970    \\
   Case 2 & 4 & 20/60 - 20/100 & Yes   & 1928    \\
   Case 3 & 6 & 20/100  & Yes  & 3547    \\
  \specialrule{1.3pt}{1pt}{1pt}
\end{tabular}

\end{adjustbox}
\end{center}
\caption{Group characterization summary.}\label{pupildata}
\end{table}

In this summary of data, N refers to the number of subjects for each group. Visual acuity indicates the range of visual acuity scores assessed by ETDRS of the better eye and AMD represents the presence (Yes) or absence (No) of AMD. Then data are classified into 4 groups based on the visual acuity and the presence or absence of AMD. The visual acuity is related to an ability to resolve fine visual detail and can be measured by the protocol outlined in the Early Treatment of Diabetic Retinopathy Study (ETDRS) \citep{Moloney2006}, which means that the group of case 3 is the worst case and the group of case 1 is the weakest among the three patient groups. Data on pupil diameter are recorded in the system at a rate of 60 HZ, or 60 times per second and a scaling factor is applied the relative recorded pupil diameter to account for camera distortion of size.

\begin{figure}[!ht]
\centering
\includegraphics [scale=0.4 , clip]{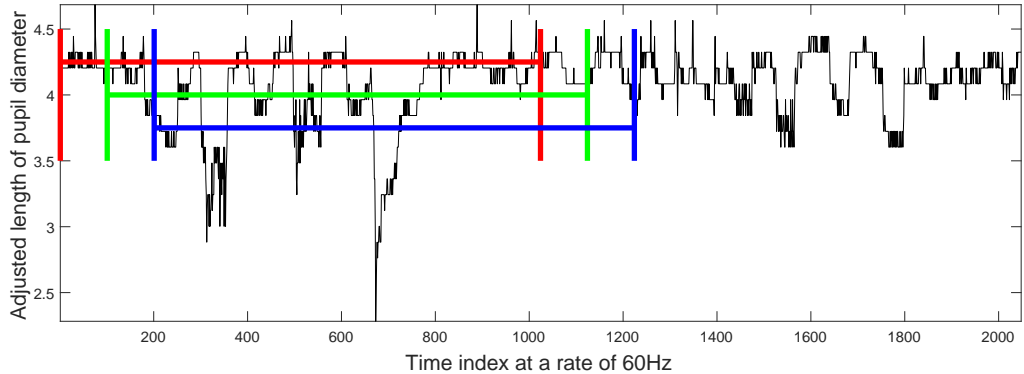}
\caption{An example of 2048 length pupillary signal segmentation. The red, green, and blue intervals represent the 1st, 2nd, and 3rd segments.}
\label{expupilsegment}
\end{figure}

Note that we segmented the signals for each individual since the number of subjects is too small due to difficulty of collecting the measurements. Another reason for the segmentation is that their lengths are not equally long. For each signal, we cut the total signal into 1024-length pieces with 100 window size. For example, we obtain total 11 dataset (segments) of 1024 length from a 2048 length pupillary signal and its visual representation is provided in Figure \ref{expupilsegment}. Table \ref{pupildata} summarizes the finalized dataset according to this segmentation concept and finally the total number of samples is 8615.

\subsubsection{Classification}{\label{sec-classification1}}
In this section we describe a way of classifying the pupillary signals based on the proposed $\text{NDWT}_\text{\large{c}}$. First, we performed the proposed 1-D $\text{NDWT}_\text{\large{c}}$ to the segmented signals found in Section \ref{sec-data1} using complex Daubechies 6 tap filter. Next, we calculated a slope of wavelet spectra explained in Section \ref{sec-NDCwavespec} and averages of phases at all level $j=J_0, \dots, J-1$ defined in Equation (\ref{avgphase}) as features.

As we discussed in Section \ref{sec-data1}, segmentation of signals can increase the number of available data. However, it also induces dependence within the data for each subject. In order to quantify and remove the dependence effects within each subject, we performed a two-way nested analysis of variance (ANOVA) under the model as
\begin{equation}\label{pupilmodel}
y_{ijk} = u + \alpha_i + \beta_{j(i)} + \epsilon_{ijk}, \;\; \epsilon_{ijk} \sim N(0, \sigma^2)
\end{equation}
with standard identifiability constraints $\sum_{i} \alpha_i = 0, \; \sum_{j} \beta_{j(i)} = 0$.
For the model (\ref{pupilmodel}), let us consider $y_{ijk}$ as the spectral slope obtained by the $\text{NDWT}_\text{\large{c}}$ for each segmented pupillary signal, then it can be decomposed to a grand mean $u$, an effect of groups on the slope $\alpha_i, \; i=1,2,3,4$, an effect of subjects on the slope $\beta_{j(1)}, \; j=1,2,\dots,6$, $\beta_{j(2)}, \; j=1,2,\dots,8$, $\beta_{j(3)}, \; j=1,2,\dots,4$, and $\beta_{j(4)}, \; j=1,2,\dots,6$ for the control, case 1, case 2, and case 3, respectively, and finally an error $\epsilon_{ijk}$. The result of the two-way nested ANOVA test based on the model (\ref{pupilmodel}) is presented in the Table \ref{pupilANOVA}.

\begin{figure}[h!tb]
  \centering
  \subfigure[]{\includegraphics[width=2.5in, height=1.5in]{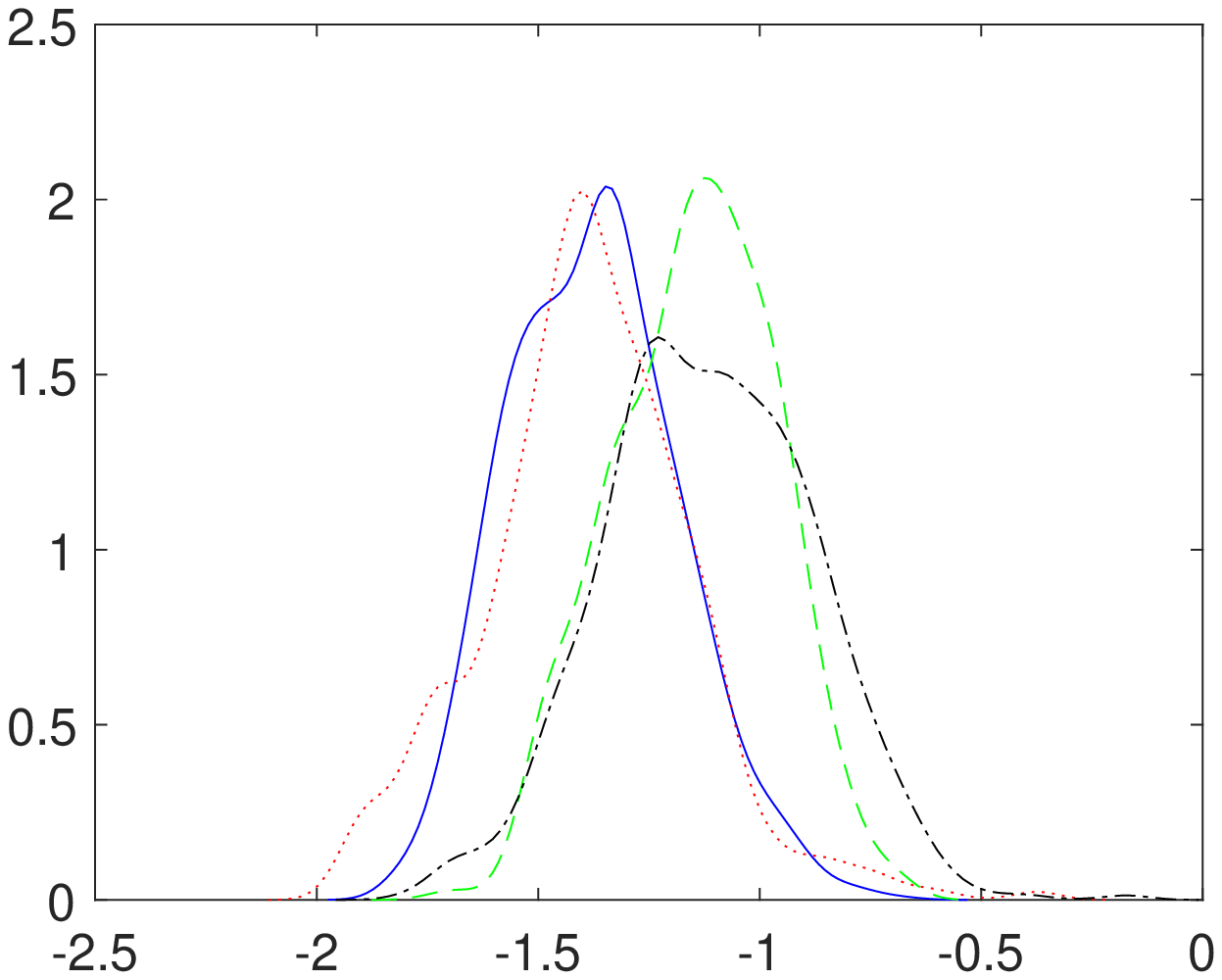}} \qquad
  \subfigure[]{\includegraphics[width=2.5in, height=1.5in]{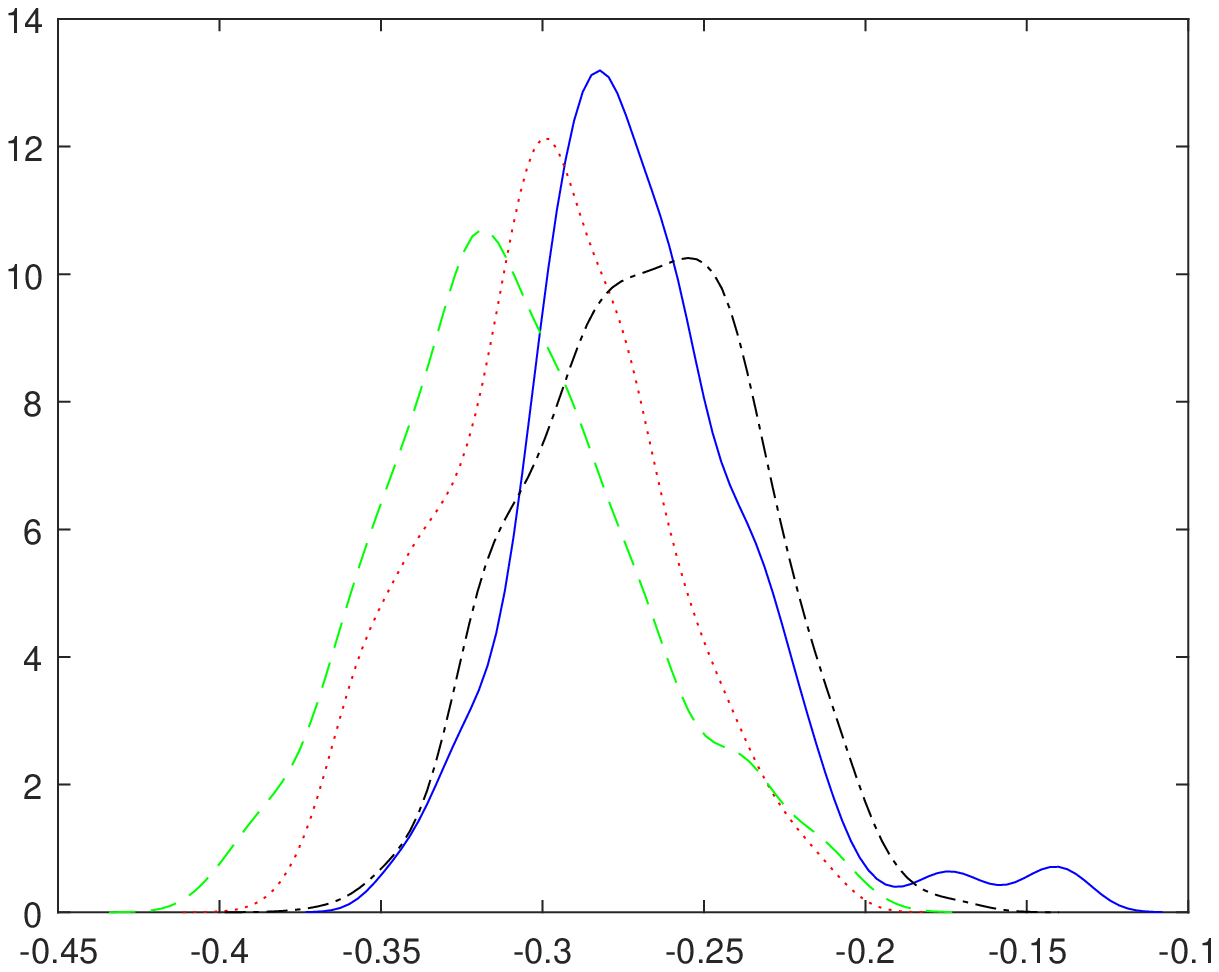}} \\
  \subfigure[]{\includegraphics[width=2.5in, height=1.5in]{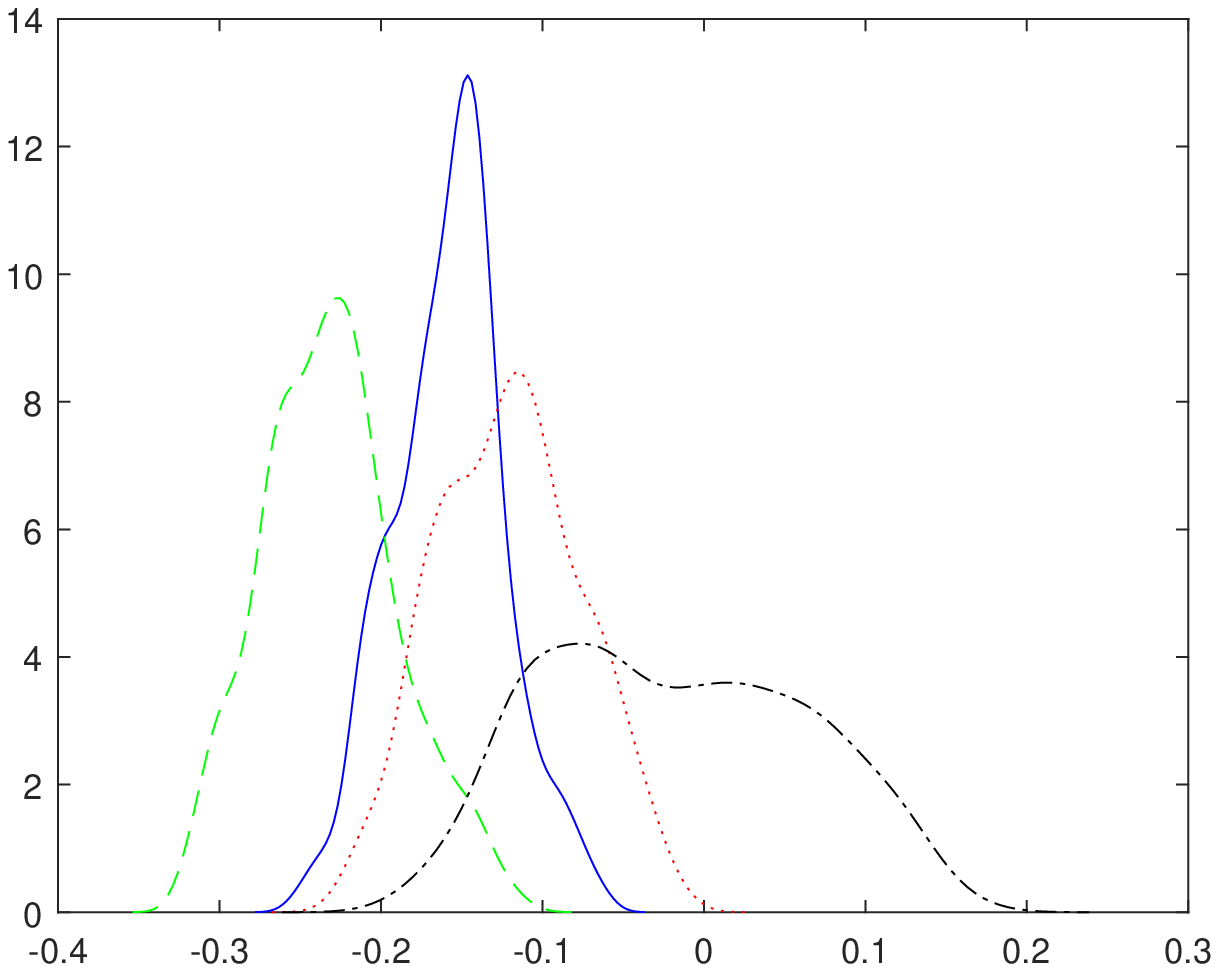}} \qquad
  \subfigure[]{\includegraphics[width=2.5in, height=1.5in]{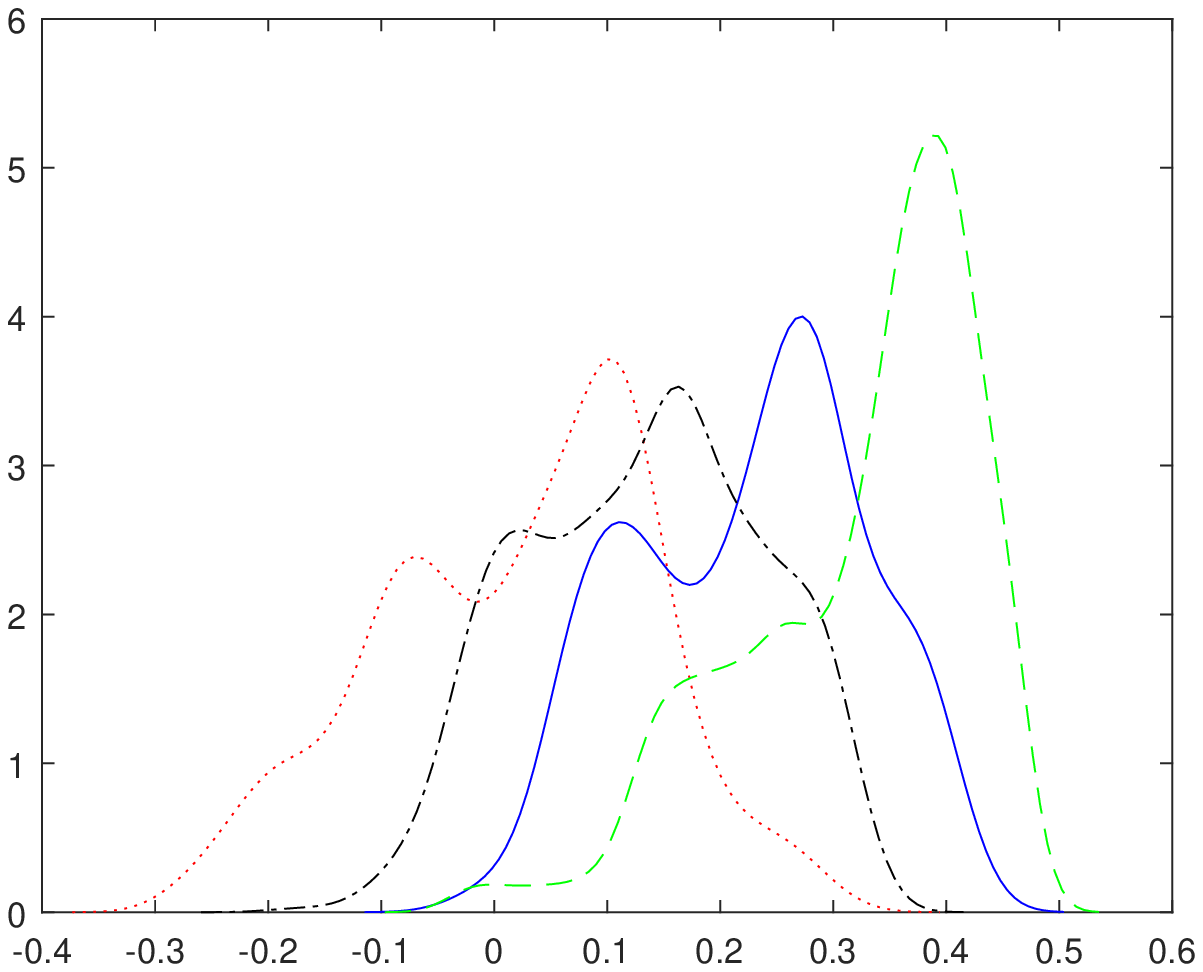}} \\

%  \subfigure[]{\includegraphics[width=2.5in, height=2.5in]{pupilphase1.eps}} \qquad
  \caption{Estimated density plots of slope in (a) and averages of phases at the last three finest levels in (b) $j=J-3$, (c) $j=J-2$, (d) $j=J-1$. The blue solid line for control, the red dotted line for case 1, the green dashed line for case 2, and the black dash-dotted line for case 3.}
  \label{fig:phase1}
\end{figure}

\begin{figure}[h!tb]
  \centering
  \subfigure[]{\includegraphics[width=2.5in, height=1.5in]{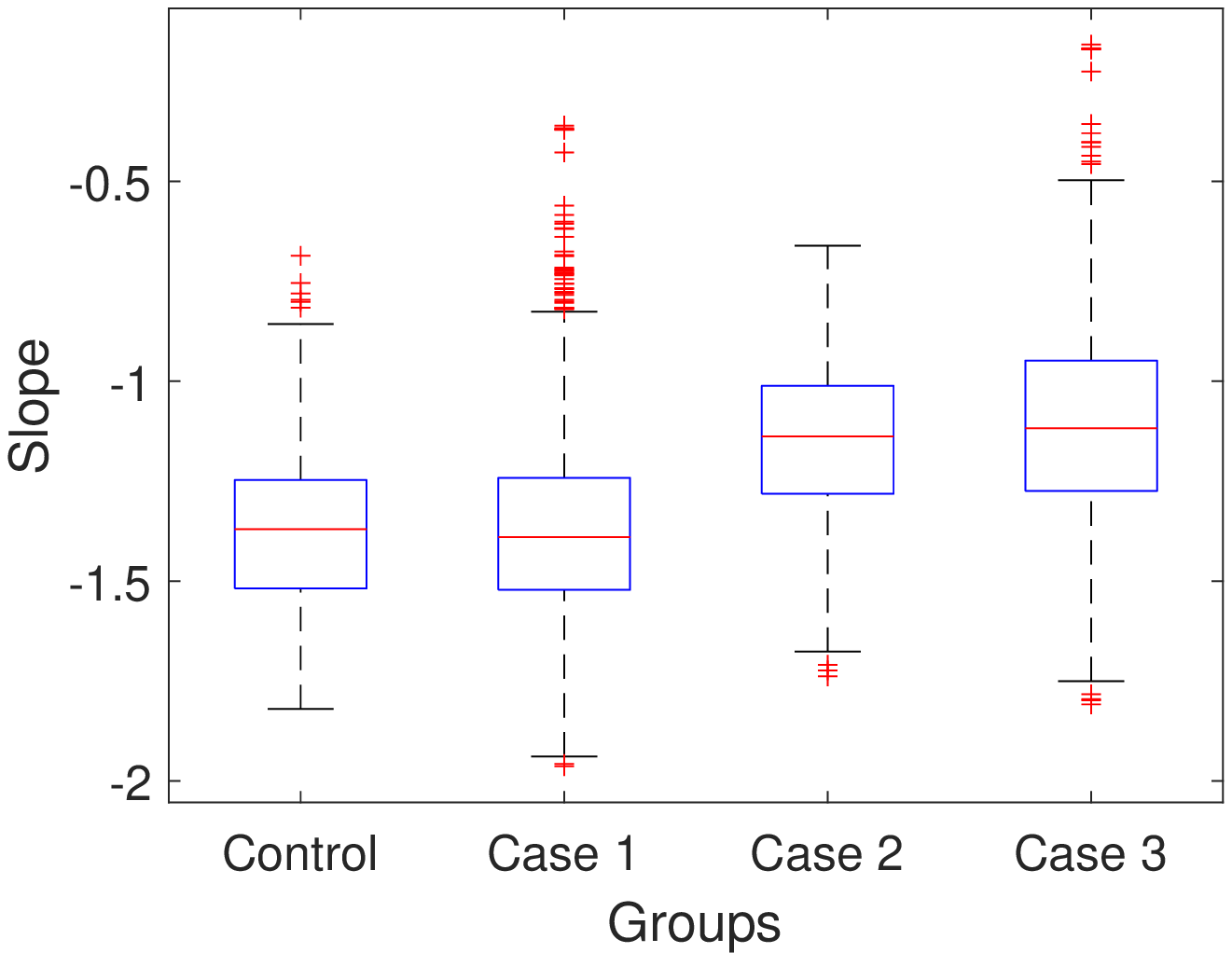}} \qquad
  \subfigure[]{\includegraphics[width=2.5in, height=1.5in]{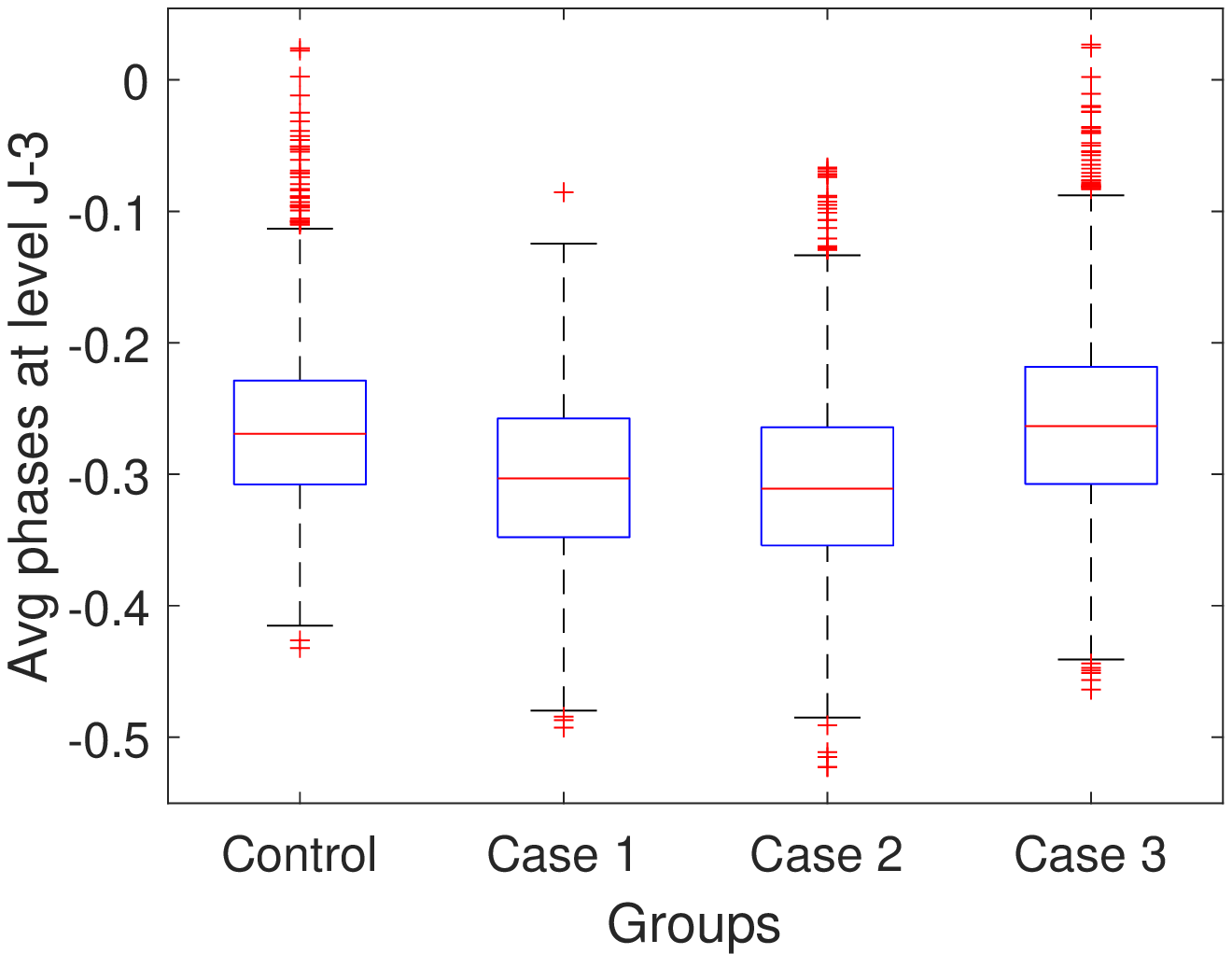}} \\
  \subfigure[]{\includegraphics[width=2.5in, height=1.5in]{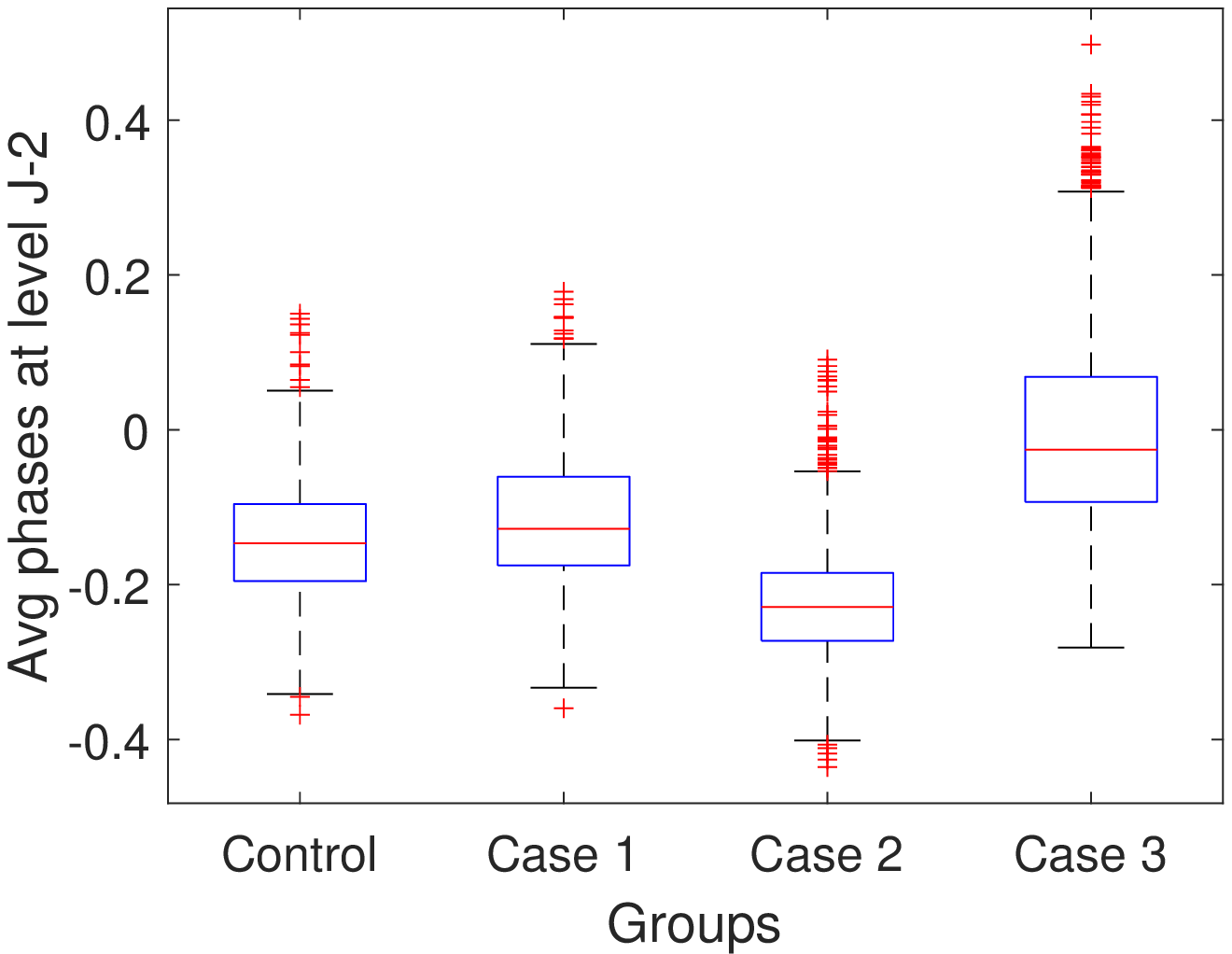}} \qquad
  \subfigure[]{\includegraphics[width=2.5in, height=1.5in]{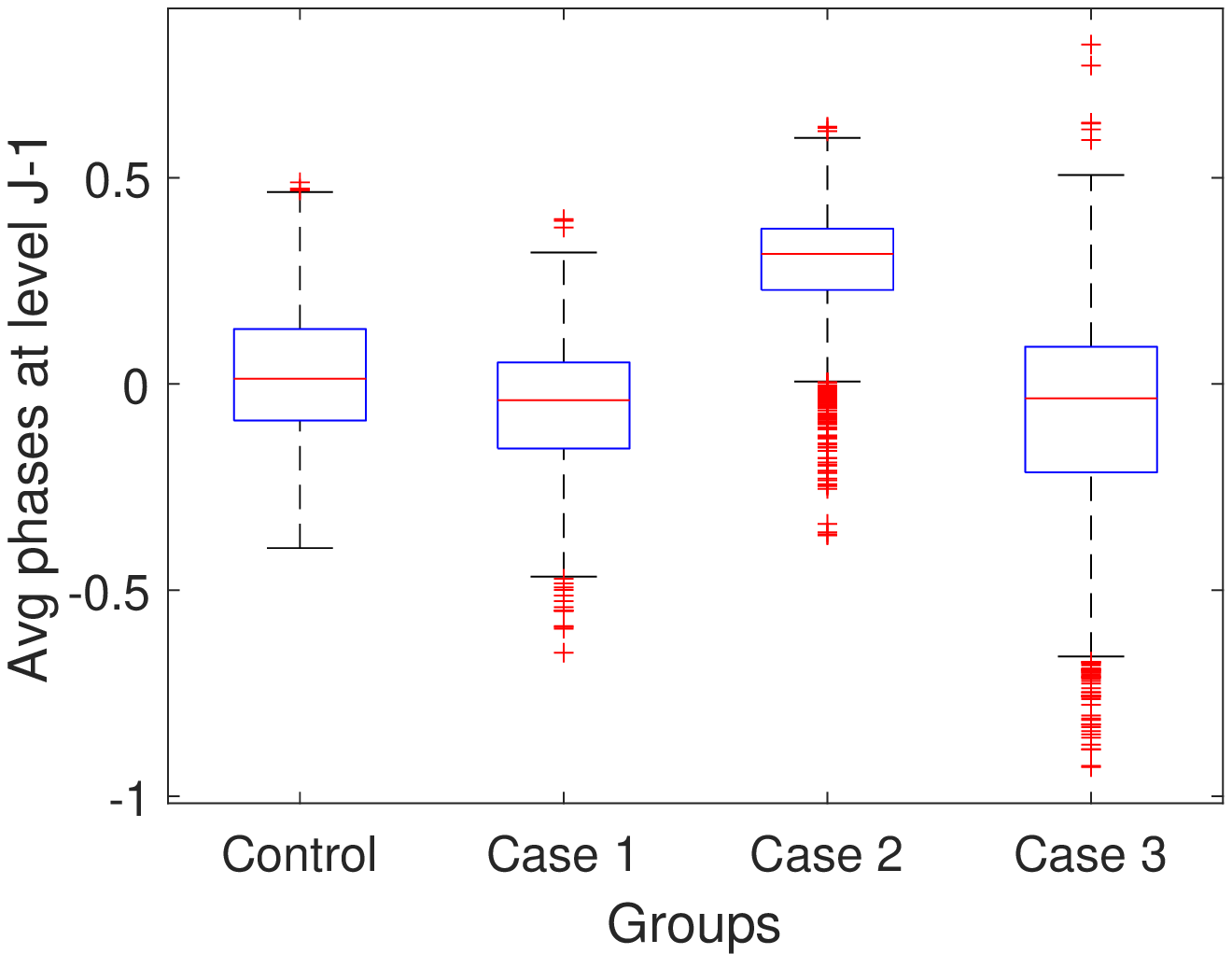}} \\

%  \subfigure[]{\includegraphics[width=2.5in, height=2.5in]{pupilphase1.eps}} \qquad
  \caption{Box plots of slope in (a) and averages of phases at the last three finest levels in (b) $j=J-3$, (c) $j=J-2$, (d) $j=J-1$.}
  \label{boxfig:phase1}
\end{figure}

\begin{table}[h!tb]
\begin{center}
\begin{adjustbox}{max width=\textwidth}

\begin{tabular}{c|c|c|c|c|c}
  \specialrule{1.3pt}{1pt}{1pt}
   Source & SSE & df & MSE & F stat & Prob$>$F  \\\hline \hline
   Group & 131.5808 & 3  &  43.8603  & 498.0589 & 0   \\
   Nested subject & 355.0408 & 20 &  17.7520 & 201.5848 & 0   \\
   Error & 756.6321 & 8592 & 0.0881   &  &   \\
   Total & 1243.2537 & 8614  &   &   &  \\
  \specialrule{1.3pt}{1pt}{1pt}
\end{tabular}

\end{adjustbox}
\end{center}
\caption{The result of the two-way nested ANOVA based on the model (\ref{pupilmodel}).}\label{pupilANOVA}
\end{table}
We can see that effects of both the groups and subjects are significantly different; the two hypotheses, $H_0 : \alpha_i = 0 \; \mbox{for all} \; i$ and $H_0 : \beta_{j(i)} = 0 \; \mbox{for all} \; i \; \mbox{and} \; j$, are rejected.
Since we are not interested in the effects of nested subjects, to represent each pupillary signal we use $y_{ijk}^* = y_{ijk} - \hat{\beta}_{j(i)}$ instead of $y_{ijk}$ for our classification analysis where $\hat{\beta}_{j(i)} = \bar{y}_{ij.} - \bar{y}_{i..}$.
All other factors such as phase averages at each level and spectral slopes from different wavelet transform methods were tested in the same way.
Every test showed comparable results with the case of the spectral slope obtained by the $\text{NDWT}_\text{\large{c}}$.
We use the $y_{ijk}^* = y_{ijk} - \hat{\beta}_{j(i)}$ instead of $y_{ijk}$  for all variables.
Estimated density plots of the slope and the three finest levels $j= \{J-3, J-2, J-1 \}$ are shown in Figure \ref{fig:phase1} and corresponding box plots in Figure \ref{boxfig:phase1}.

Using the two types of extracted descriptors with such modifications, we employed gradient boosting to classify the pupillary signals. We also considered random forest, k-NN, and SVM, however, the gradient boosting consistently outperformed the rest. For simulations, we randomly split the dataset to training and testing sets in proportion 75\% to 25\%, respectively. This random partition to training and testing sets was repeated $1,000$ times, and the reported prediction measures are averages over the $1,000$ runs.

\subsubsection{Results}{\label{sec-result1}}
Since there are four labeled groups, we evaluated performances of the suggested $\text{NDWT}_\text{\large{c}}$ in the context of overall accuracy and sensitivities of the four groups as shown in Table \ref{pupiltable}.
For comparisons, we also performed the standard WT and NDWT using Haar filter, and $\text{WT}_\text{\large{c}}$ using the same complex Daubechies 6 tap filter of the $\text{NDWT}_\text{\large{c}}$.

\begin{table}[h!tb]
\begin{center}
\begin{adjustbox}{max width=\textwidth}

\begin{tabular}{c|c|c|c|cccc|c}
  \specialrule{1.3pt}{1pt}{1pt}
   Order & Transform & Features & \begin{tabular}{@{}c@{}} Overall \\ Accuracy rate\end{tabular}
 & \begin{tabular}{@{}c@{}}Sensitivity \\ Control\end{tabular}
 & \begin{tabular}{@{}c@{}}Sensitivity \\ Case 1\end{tabular} & \begin{tabular}{@{}c@{}}Sensitivity \\ Case 2\end{tabular} & \begin{tabular}{@{}c@{}}Sensitivity \\ Case 3\end{tabular} & \begin{tabular}{@{}c@{}}Computing \\ Time  \end{tabular} \\\hline \hline
   $1$st & WT & Slope  &  0.4458	& 0.0913	& 0.4119	& 0.0513	& 0.7966 & 0.0150s \\ \hline
  $2$nd & $\text{WT}_\text{\large{c}}$ & Slope  &  0.3992	& 0.2066 &	0.3018	& 0.1265	& 0.6668 & \\
   $3$rd & & $\angle d_j$ &  0.5808 &	0.2117 &	0.2755	& 0.8395 &	0.7332 & 0.0207s \\
   $4$th & &  Slope + $\angle d_j$ & 0.6685	& 0.3301	& 0.3728	& 0.8729 &	0.8338 & \\ \hline
  $5$th & NDWT & Slope  & 0.4596	& 0.0916	& 0.4799	& 0.0635	& 0.7856 & 0.0192s \\ \hline
  $6$th & $\text{NDWT}_\text{\large{c}}$ & Slope  & 0.4172	&0.1762	&0.3936	&0.2178	&0.6184 & \\
  $7$th &  & $\angle d_j$ & 0.7753 &	0.6588 &	0.6235 &	0.8755	& 0.8431  & 0.0270s \\
  $8$th & & Slope + $\angle d_j$ & \textbf{0.8226} &	\textbf{0.6857}	& \textbf{0.7263}	& \textbf{0.8864}	& \textbf{0.8870} & \\
  \specialrule{1.3pt}{1pt}{1pt}
\end{tabular}

\end{adjustbox}
\end{center}
\caption{Gradient boosting classification results. Total 8 methods are compared and the best result is obtained by the $\text{NDWT}_\text{\large{c}}$ with slopes and averages of phases.}\label{pupiltable}
\end{table}

For the convenience, we named the methods in order from $1$ to $8$.
Note that for the $\text{WT}_\text{\large{c}}$ and $\text{NDWT}_\text{\large{c}}$, the phase averages are more discriminatory than the slopes and combinations of the two gives the best results.
Another interesting finding is that classifiers without phase information tend to show low performance in classifying the control and case 2.
In this context, we can conclude that information that separates the control and case 2 is located in the phase.
Comparing the $4$th and $5$th to the $8$th, one can see that the $\text{NDWT}_\text{\large{c}}$ dominates both $\text{WT}_\text{\large{c}}$ and NDWT.
Therefore, the performance improves if the wavelet spectra from $\text{NDWT}_\text{\large{c}}$ with additional descriptors based on phase are used.

Finally, we recorded calculational complexity (in terms of times) for all considered versions of wavelets (WT, $\text{WT}_\text{\large{c}}$, NDWT, $\text{NDWT}_\text{\large{c}}$) to transform one 1024-length signal.
As expected, the computation times are proportion to the overall accuracies: more accurate results take longer to calculate.

\subsection{Application in Screening Mammograms}{\label{sec-mammo}}
Breast cancer is the second leading cause of cancer-related death in women in the United States. The National Cancer Institute's research in \citet{Altekruse2010} estimated that 1 in 8 women is likely to develop breast cancer during their lives. The U.S Department of Health and Human Services set a goal to reduce  breast cancer death rate by 10\% by 2020.
Mammography is the one of the widely-used screening methods for early detection of breast cancer which can improve prognosis as well as lead to less invasive interventions \citep{NCI2014}. However, the radiological interpretation of mammogram images is a difficult task due to the heterogeneous nature of normal breast tissue. In other words, it is difficult to classify cancerous and non-cancerous images by
merely looking at them. Moreover, cancers can be of similar radiographic density as normal tissue, which can affect correct detection and decrease the sensitivity of the tests. Specificity of detection is of concern as well because it was observed that of the 5\% of the mammogram images suggested for further testing, as many as 93\% turned out to be false positives \citep{Houssami2006}. Therefore,  it is very important to improve both the sensitivity and specificity of the mammographic diagnostics.

It is well-known fact that one of the testing modalities is a density and fine scale structure of
the breast tissue.
This indicates that the scaling information of the digitized images can be utilized for classification.
Some previous work on mammogram classification by using a wavelet spectra can be found in \citet{Jeon2014},  \citet{Roberts2017}, and \citet{Feng2018}.
Since the wavelet spectra captures information contained in the background tissue of images rather than predefined templates of expected cancer morphology (tumors and microcalcifications), the spectral descriptors provide for a new and independent modality for diagnostic testing.

A study of \citet{Jeon2014} suggested a classification procedure based on the estimated slope of modulus and phase average from the finest level in a $\text{WT}_\text{\large{c}}$ transformed image.
As mentioned in the previous section,   the method showed  relatively low classification accuracy in spite of better balancing specificity and sensitivity compared to other wavelet-based methods using real-valued wavelets. Another disadvantage of the method was that it only can be applied to squared images of dyadic size, since it is based on the standard WT.
In studies by \citet{Jeon2014}, and \citet{Roberts2017} the mammogram images were manually split into 5 dyadic sub-images due to this limitation in experiments.
This manual selection of sub-images is impractical for screening mammogram images and even causes a problem of multicollinearity due to overlapping.
The study of \citet{Kang2019} resolved these problems by using the NDWT with non-decimation property. However, its classification results can be improved more if the non-decimated complex wavelet transform is used.
In the next section, we provide classification results using all fore-mentioned methods including the proposed $\text{NDWT}_\text{\large{c}}$ and demonstrate that the latter dominates others.

\subsubsection{Description of Data}
\begin{figure}[!ht]
\centering
\includegraphics[scale=0.8]{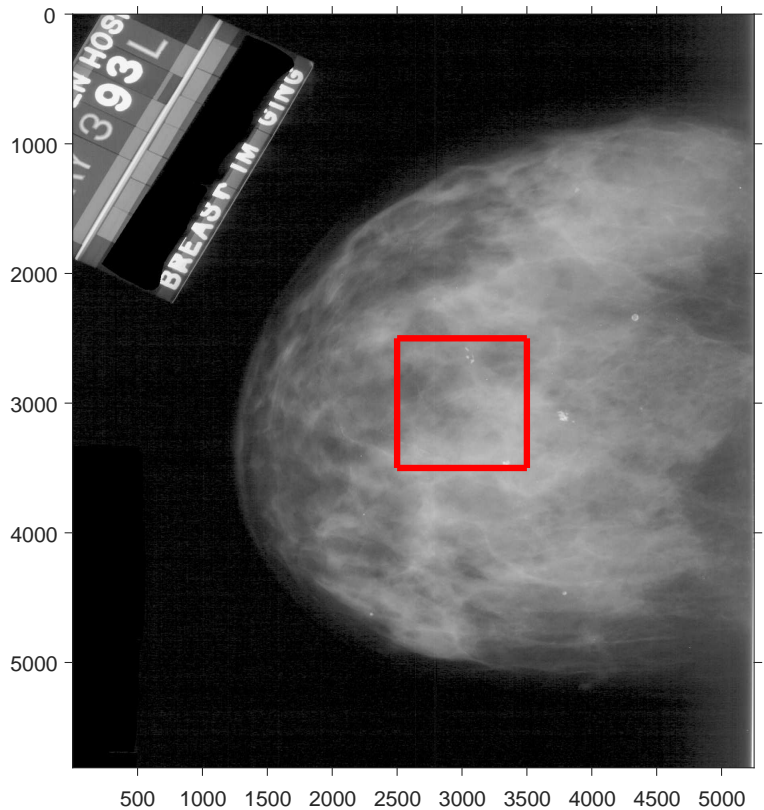}
\caption{An example of mammogram image. The $1024 \times 1024$ area surrounded by red lines indicates the ROI.}
\label{ROI}
\end{figure}

{\label{sec-data}}
The collection of digitized mammograms for analysis was obtained from the University of South Florida's Digital Database for Screening Mammography (DDSM), which are explained in detail in \citet{Heath2001}. Images from this database containing suspicious areas are accompanied gold standard true label assessed and verified through biopsy. We selected 45 normal controls (benign) and 79 cancer cases (malignant) scanned on the HOWTEK scanner at the full 43.5 micron per pixel spatial resolution. Each case contains craniocaudal (CC) and mediolateral oblique (MLO) projection mammograms from a screening exam. We only analyze the CC projections.  Note that an image containing an area outside of breast can seriously impact the result when self-similarity features are used in classification.
Since the outside area is smooth the spectral slope may appear steeper. To resolve this problem, the studies in \citet{Jeon2014}, and \citet{Roberts2017} split the mammogram images into 5 sub-images within the tissue region. This
image-by-image splitting method, however, has problems since some subimages partially overlapped.
Instead, the study in \citet{Kang2019} used a mask-based method to remove irrelevant parts of the mammogram image, and to define self-similarity properties based on coefficients belonging only to tissue part.
However, the masked images also covered the side regions of breast tissues that are unlikely to contain significant information on the cancer status. Thus, we alternatively select a single region of interest (ROI) from each mammogram image as illustrated in Figure \ref{ROI}. Even though we could analyze images of any size, thanks to the non-decimation property, the sub-images of size $1024 \times 1024$ were manually selected because some other methods used for comparisons require dyadic image dimensions.

\subsubsection{Classification}
\begin{figure}[h!tb]
  \centering
  \subfigure[]{\includegraphics[width=2.5in, height=1.5in]{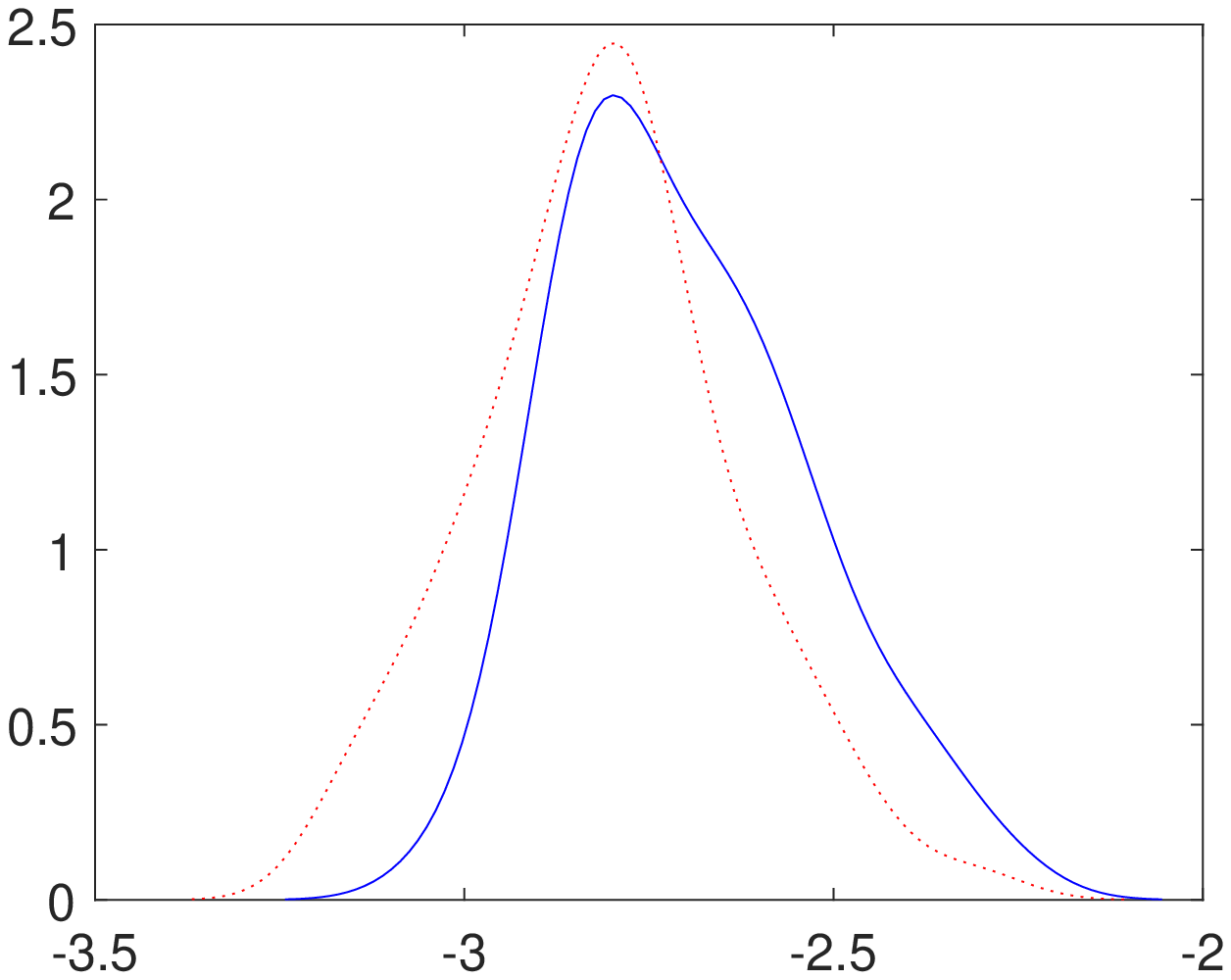}} \qquad
  \subfigure[]{\includegraphics[width=2.5in, height=1.5in]{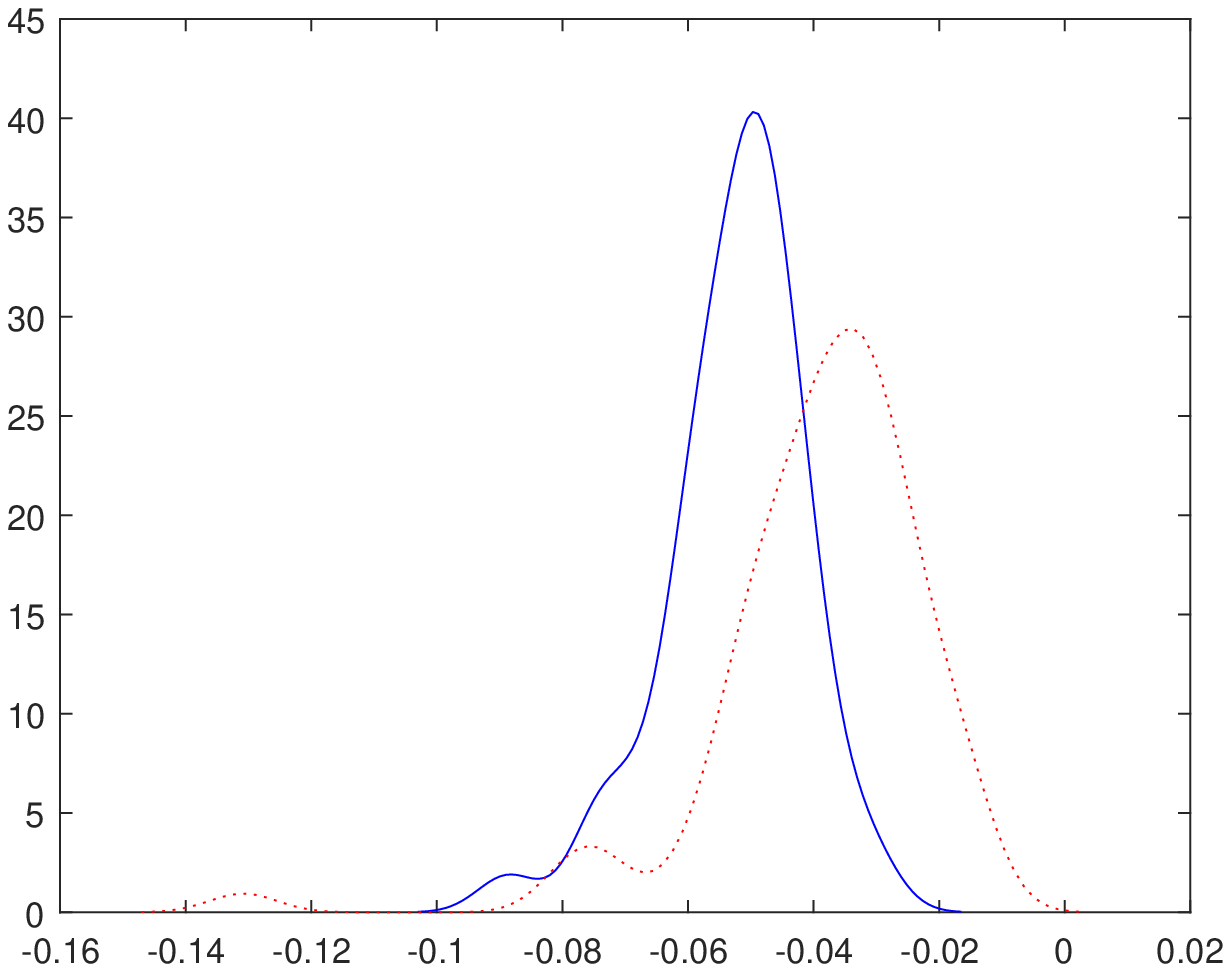}} \\
  \subfigure[]{\includegraphics[width=2.5in, height=1.5in]{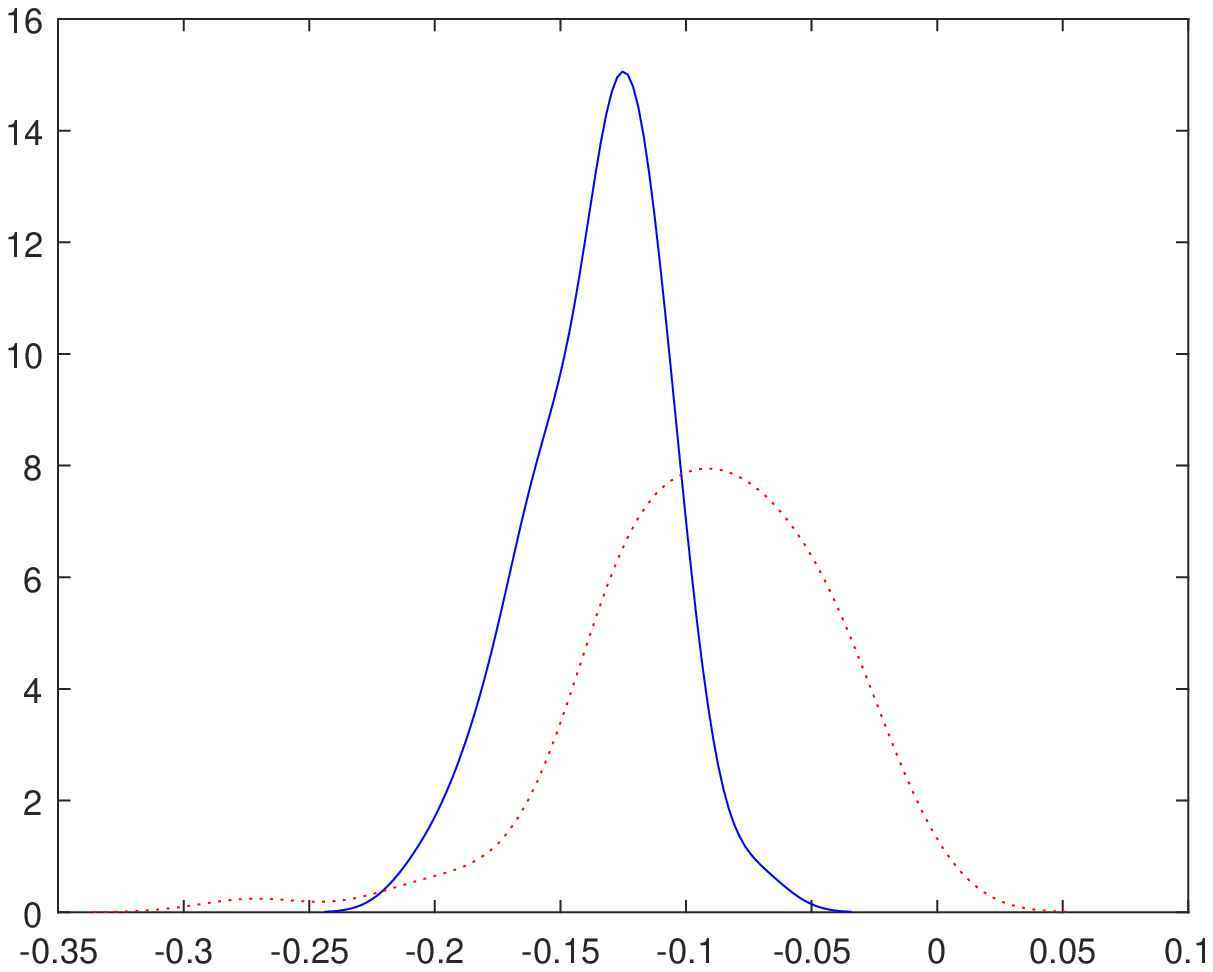}} \qquad
  \subfigure[]{\includegraphics[width=2.5in, height=1.5in]{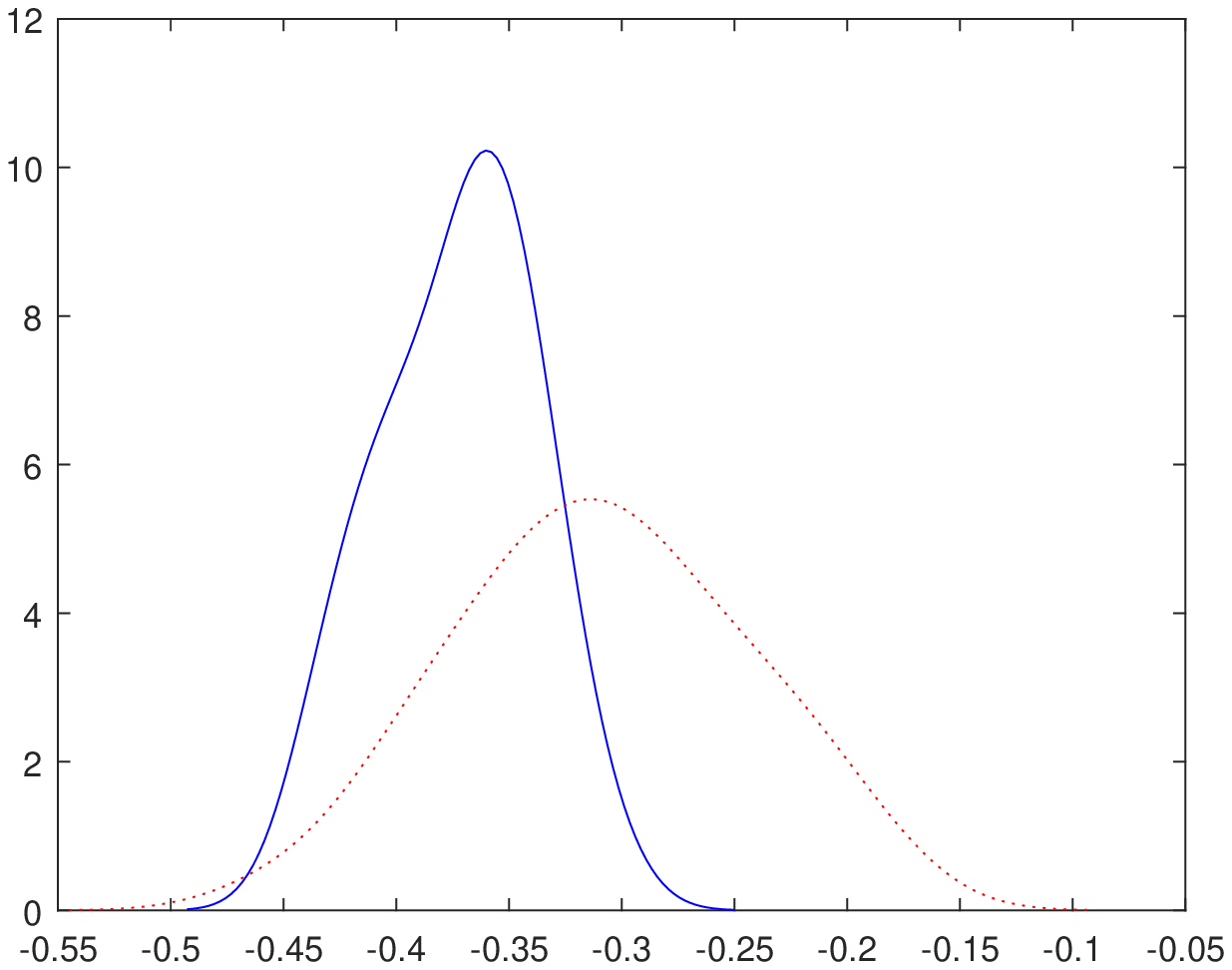}} \\
  \caption{Empirical density plots of slope in (a) and phase averages at the three finest levels in (b) $j=J-3$, (c) $j=J-2$, (d) $j=J-1$. The blue solid line is for normal controls while the red dotted line is for cancer cases.}
  \label{fig:phase}
\end{figure}

\begin{figure}[h!tb]
  \centering
  \subfigure[]{\includegraphics[width=2.5in, height=1.5in]{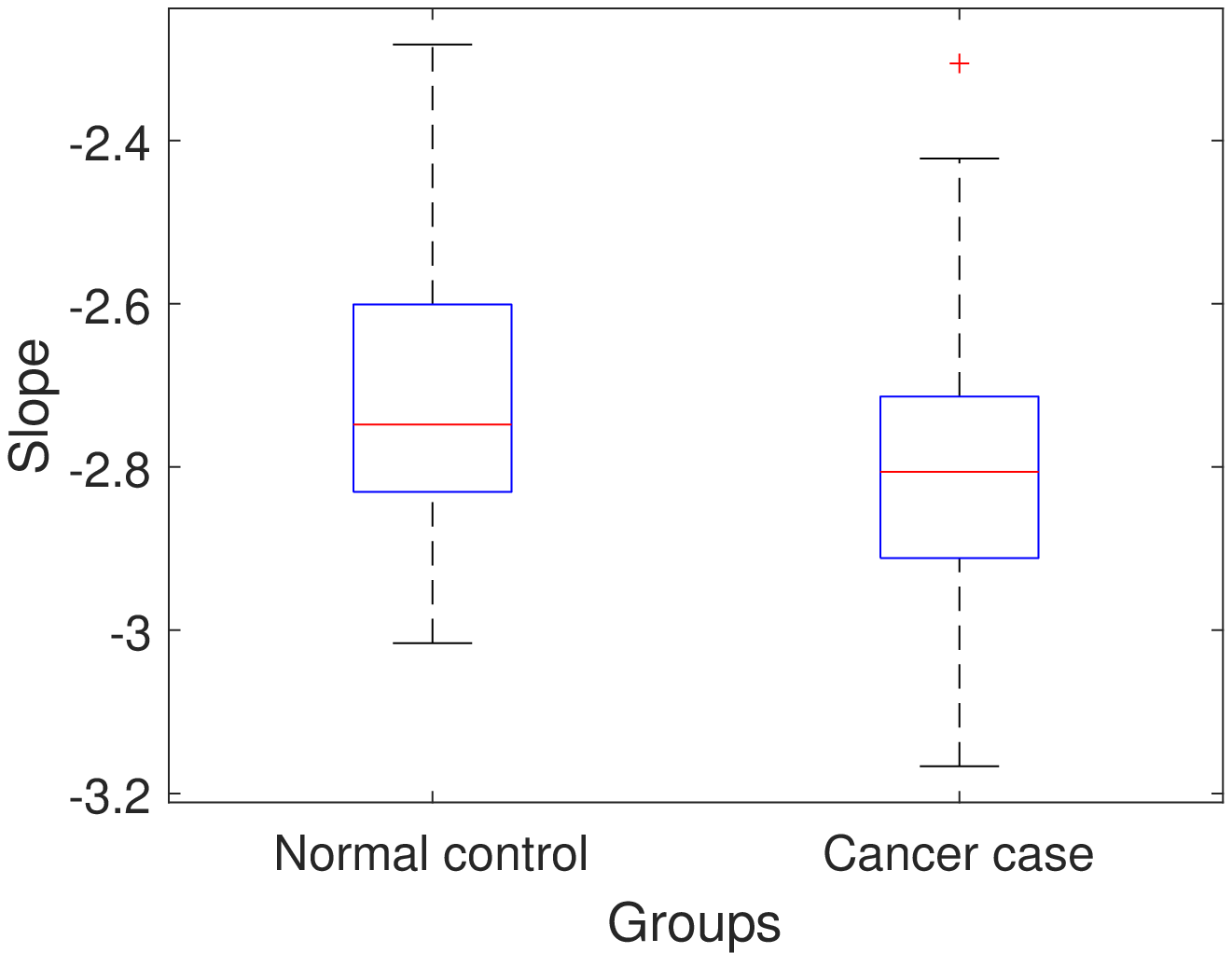}} \qquad
  \subfigure[]{\includegraphics[width=2.5in, height=1.5in]{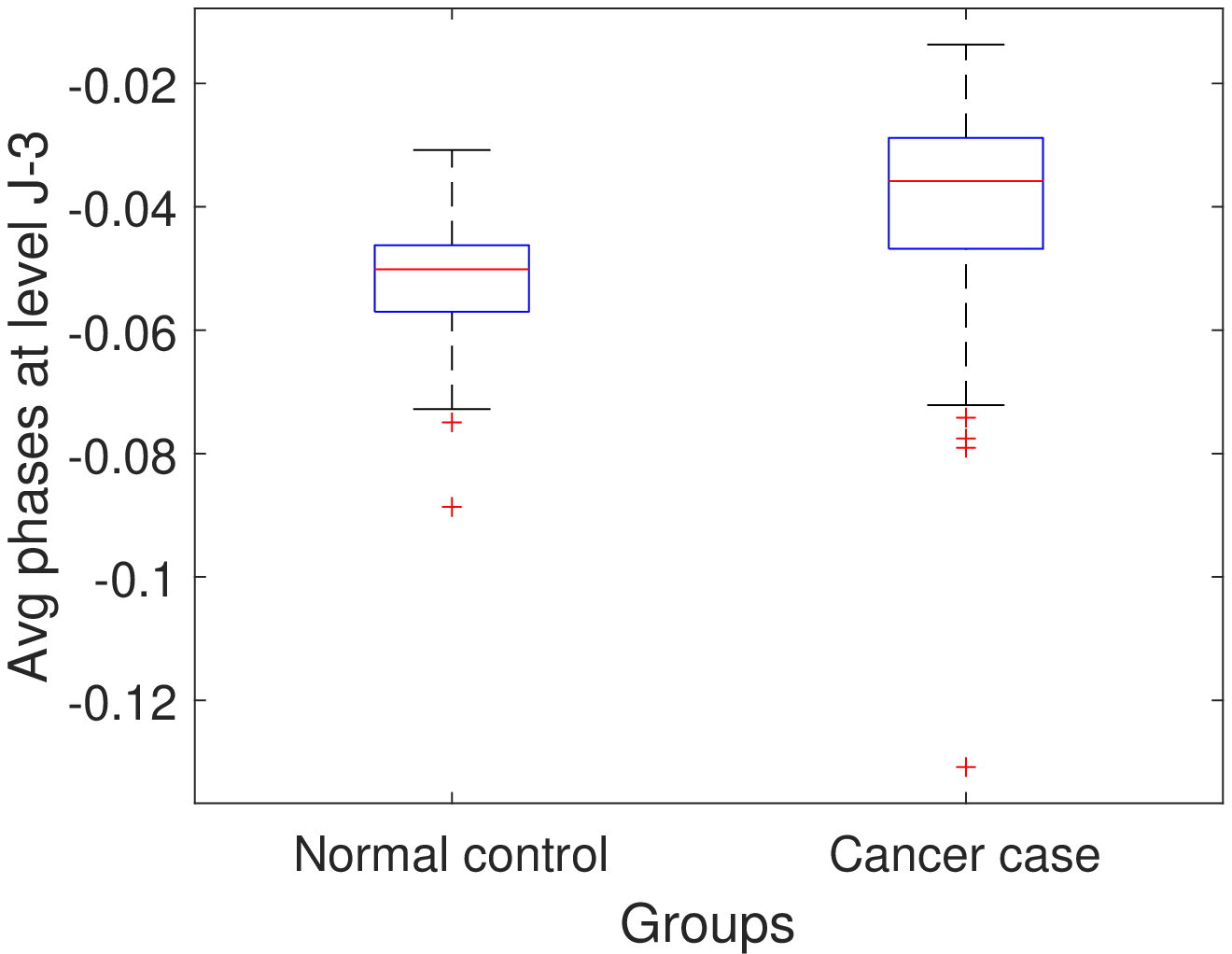}} \\
  \subfigure[]{\includegraphics[width=2.5in, height=1.5in]{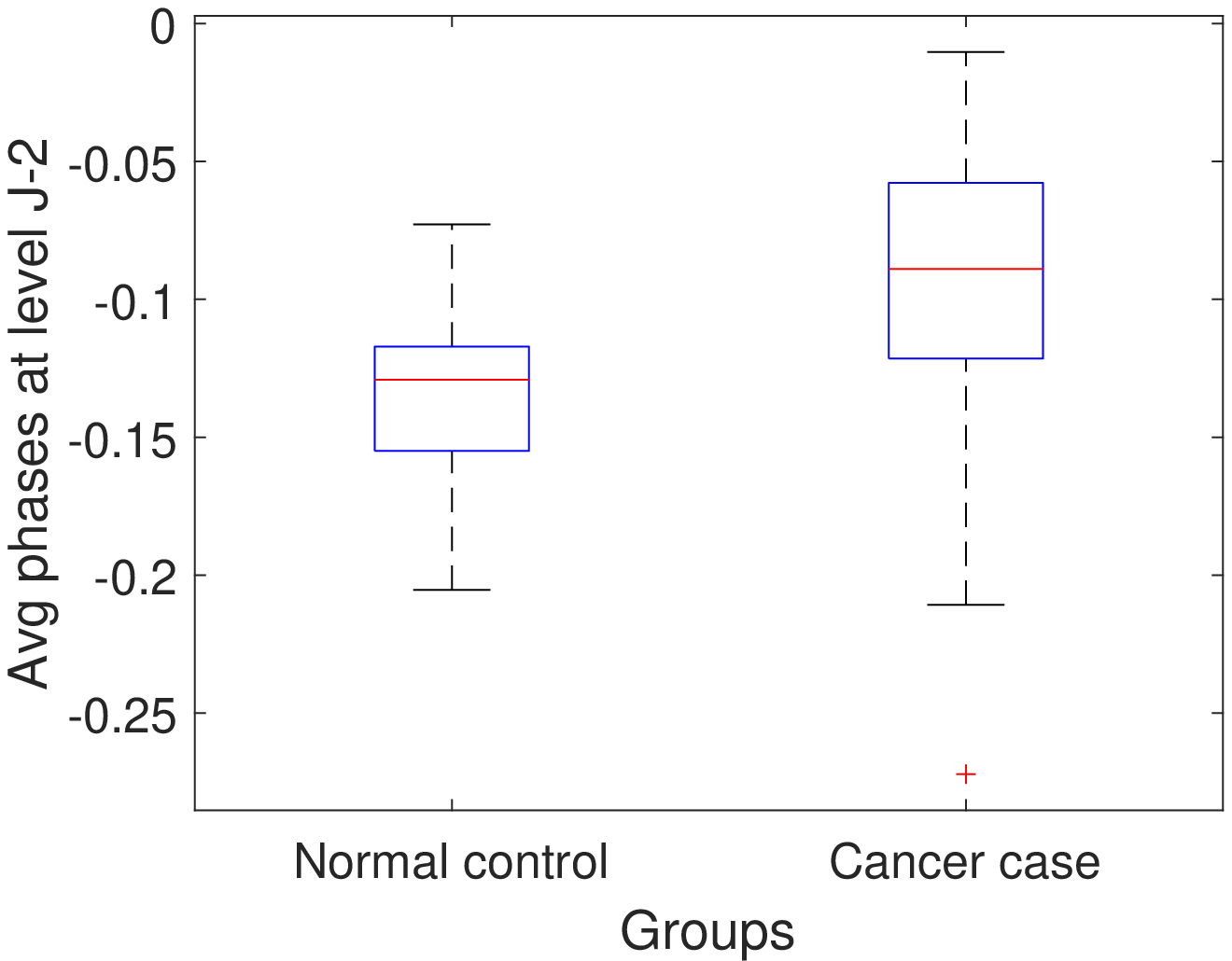}} \qquad
  \subfigure[]{\includegraphics[width=2.5in, height=1.5in]{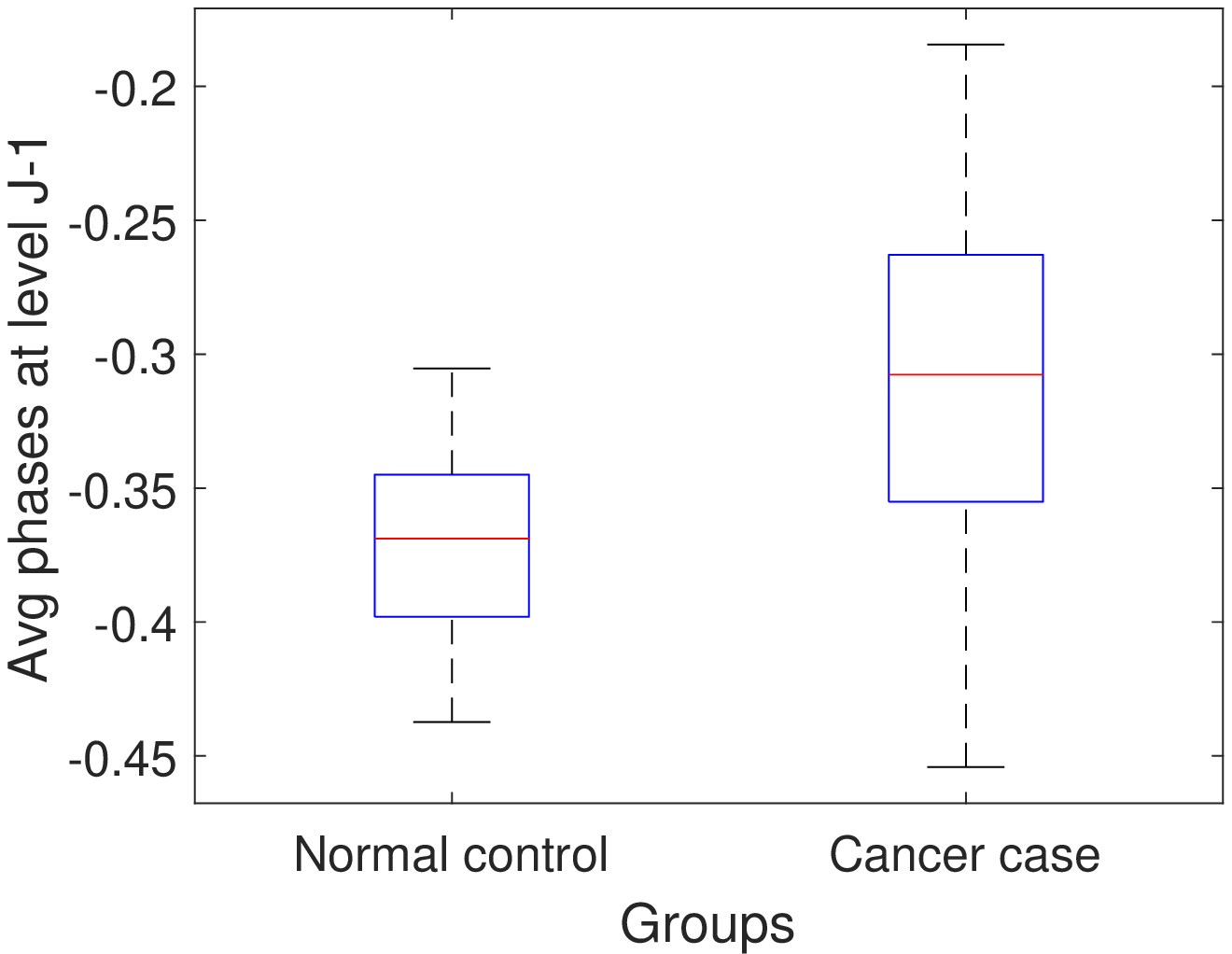}} \\
  \caption{Box plots of slope in (a) and phase averages at the last three finest levels in (b) $j=J-3$, (c) $j=J-2$, (d) $j=J-1$.}
  \label{boxfig:phase}
\end{figure}

{\label{sec-classification}}
In this section we explain how we classified the mammogram images. First, on the ROI images from Section \ref{sec-data} we applied the scale-mixing 2-D $\text{NDWT}_\text{\large{c}}$ with $s=0$ as well as WT, NDWT, and $\text{WT}_\text{\large{c}}$  for comparison.
Next, we calculated spectral slope described in Section \ref{sec-scNDCwavespec} and phase averages for all level $j=J_0, \dots, J-1$ defined in Equation (\ref{avgphase}). These were features that were inputs to classification analysis. Empirical density plots of the slope and the last three finest levels $j= \{J-3, J-2, J-1 \}$ are displayed in Figure \ref{fig:phase} with corresponding box plots displayed in Figure \ref{boxfig:phase}.
It is evident that the differences between the classes are more pronounced in the phase-based features than the spectral slopes.

Last, we employed random forest to classify the mammogram images. We also considered the logistic regression, k-NN, SVM, and gradient boosting, however, the random forest consistently outperformed the competitors. Since the dataset is imbalanced and has a relatively small size, we selected 75\% for training and 25\% for testing at random for both control and case samples. The classification was repeated 1,000 times, and the prediction measures were obtained by averaging over 1,000 runs.

\subsubsection{Results}{\label{sec-result}}
We compared classification performances in the context of sensitivity, specificity, and overall accuracy rate, which are shown in Table \ref{mammotable}. Haar filter was chosen for WT and NDWT. To simplify the notations, we denote the phase average as $\angle d_{j}$ instead of $\angle d_{j,j}$.

\begin{table}[h!tb]
\begin{center}
\begin{adjustbox}{max width=\textwidth}

\begin{tabular}{c|c|c|ccc|c}
  \specialrule{1.3pt}{1pt}{1pt}
   Order & Transform & Features & Accuracy rate & Specificity & Sensitivity & Computing Time \\\hline \hline
   $1$st & WT & Slope  &  0.5306 &	0.3651&	0.6302  & 0.0724s \\\hline
  $2$nd & $\text{WT}_\text{\large{c}}$ & Slope  & 0.4900&	0.3114	&0.5991    & \\
   $3$rd & & $\angle d_{j}$ &  0.5117	&0.3347&	0.6175&  0.3378s \\
   $4$th & &  Slope + $\angle d_{j}$ & 0.5173	&0.2975	&0.6505 & \\\hline
  $5$th & NDWT & Slope  &  0.5571	&0.3726&	0.6694  & 2.3428s \\\hline
  $6$th & $\text{NDWT}_\text{\large{c}}$ & Slope  &  0.5453	&0.3667&	0.6541 & \\
  $7$th & & $\angle d_{j}$  & \textbf{0.7456}&	\textbf{0.7038}	&\textbf{0.7748} & 8.2451s \\
  $8$th & & Slope + $\angle d_{j}$ &  0.7342	&0.6954	&0.7617 & \\
  \specialrule{1.3pt}{1pt}{1pt}
\end{tabular}

\end{adjustbox}
\end{center}
\caption{Random forest classification results. Total 8 methods are compared and the best result is achieved by the $\text{NDWT}_\text{\large{c}}$ with only phase-based features.}\label{mammotable}
\end{table}

For simplicity, we numbered methods used in this comparative study from $1$ to $8$.
Comparing the $4$th and $5$th to the $8$th, we can see that the $\text{NDWT}_\text{\large{c}}$ dominates both $\text{WT}_\text{\large{c}}$ and NDWT.
It is also notable that the phase averages dominate slopes when comparing $6$th with $7$th, and even the phase averages alone is slightly outperform the slope in comparing $7$th with $8$th.
This, of course, may not be the case for other data, but these results emphasized the discriminatory power of the phase information.
Note that specificity significantly increased when the phase averages of $\text{NDWT}_\text{\large{c}}$ are included.
In conclusion, we can see that the best performance is achieved by $7$th method which is based on $\text{NDWT}_\text{\large{c}}$ with only phase-based features.

Similar to the 1-D application, we recorded computation times for all considered versions of transforms (WT, $\text{WT}_\text{\large{c}}$, NDWT, $\text{NDWT}_\text{\large{c}}$) to transform one $1024 \times 1024$ image.
Here the computation times also increase with the increase of overall accuracies, as in with 1-D case, however, the rate of increase is much larger.
This is because in 2-D the wavelet transform needs double matrix multiplication, compared to single in 1-D case.
Although the times rapidly increase, they are still in a reasonable range, for $\text{NDWT}_\text{\large{c}}$ takes approximately 8 seconds per image.

In a final comparison, we applied CNN (Convolutional Neural Network) which is the state-of-art image analyzing tool.
Our goal of this additional experiment was to compare CNN with the proposed method in terms of accuracy and computing time.
Tensorflow 1.5.0 in Python 3.5.2 was used for CNN with 5 layers, 0.001 learning rate, 11 batch size, and 100 training epochs and MATLAB 9.1.0 is for $\text{NDWT}_\text{\large{c}}$ on Intel(R) Core(TM) i7-6500U CPU at 2.50GHz with 12GM RAM.
We found their computing times notably different.
For the $\text{NDWT}_\text{\large{c}}$, the time for extracting features was 17 mins and then 1000 iterations of training and testing took additional 56 sec.
Thus, the total processing time was approximately 17 mins 56 secs.
In contrast, the CNN took 15 hours 1 min on average for its one-time training and testing.
Given large size of training data, the CNN did not need multiple training because large size of testing data was also available.
However, due to a limited number of mammogram images, multiple training for validation was needed.
This would take approximately $15 \times 1000$ hours for 1000 iterations.
Worse yet, the average accuracy for 10 iterations was 0.6250 with 0.4286 specificity and 0.7059 sensitivity; these are inferior to the $\text{NDWT}_\text{\large{c}}$ counterparts.
One explanation is the following.
The information on cancerous or non-cancerous tissue is strongly related to details, which are linked to the self-similarity, as discussed before.
Generally, the CNN is well known for its superb performance on classifying MNIST or CIFAR-10 where detail information is not critical.
On the other hand, the wavelet-based classifiers are very useful when critical information is located not in the coarse approximations but details, such as noise dynamics, for example.

\section{Conclusions and Future Studies}\label{sec-Conc}
In this paper, we explored a non-decimated complex wavelet transform ($\text{NDWT}_\text{\large{c}}$) for both 1-D and 2-D cases.
We demonstrated that the proposed spectra performs well in classification problems, with phase-based statistics improving the classification accuracy.
We presented comparative simulations in two real-life applications and found that the classification procedures induced by the $\text{NDWT}_\text{\large{c}}$ outperforms the $\text{WT}_\text{\large{c}}$ and NDWT.
Thus, the $\text{NDWT}_\text{\large{c}}$ may be of interest to researchers seeking more efficient wavelet-based classification method for signals or images with intrinsic self-similarity.

As a possible future directions we may be interested in different ways of calculating the spectral slopes, as similarly as in \citet{Hamilton2011} or \citet{Feng2018}.
Additionally, for the scale-mixing 2-D $\text{NDWT}_\text{\large{c}}$, using $d^{(h)}$ and $d^{(v)}$ in addition to $d^{(d)}$ for phase statistics could potentially improve the performance.
Finally using different wavelet filters for rows and columns in the scale-mixing 2-D $\text{NDWT}_\text{\large{c}}$  would provide more modeling freedom.
For instance, one can search for a wavelet, or pair of  wavelets, in a library of complex-valued wavelets for which classification is optimal.

In the spirit of reproducible research we prepared an illustrative demo as a stand alone MATLAB software with solved examples.
The demo is posted on the repository Jacket Wavelets \url{http://gtwavelet.bme.gatech.edu/}.

\vspace*{0.4in}
\noindent
{\bf Acknowledgement.~}
We thank Seonghye Jeon and Minkyoung Kang for the mammogram data, and Bin Shi for the pupil-diameter data.
This research was in part supported by NSF grant DMS-1613258 and Giglio Family Cancer Research Award.

\vspace{20cm}

\bibliographystyle{plainnat}

%\bibliography{ComplexND_Bib}

\end{document}